\documentclass[aps,preprint,floats,nofootinbib]{revtex4}
\usepackage{amsmath}
\usepackage{amssymb}
\usepackage{latexsym}
\usepackage{epsf}
\usepackage{psfrag}
\usepackage{feynarts}
\usepackage{graphicx}

\newlength{\defbaselineskip}
\newcommand{\setlinespacing}[1]%
           {\setlength{\baselineskip}{#1 \defbaselineskip}}

\def\lsim{\mathrel{\raise.3ex\hbox{$<$\kern-.75em\lower1ex\hbox{$\sim$}}}} 
\def\gsim{\mathrel{\raise.3ex\hbox{$>$\kern-.75em\lower1ex\hbox{$\sim$}}}} 
\begin{document}

\preprint{
\hfill
\begin{minipage}[t]{3in}
\begin{flushright}
\vspace{0.0in}
FERMILAB--PUB--06--374--A--T
\end{flushright}
\end{minipage}
}

\hfill$\vcenter{\hbox{}}$

\vskip 0.5cm

\title {The Interplay Between Collider Searches For Supersymmetric Higgs Bosons and Direct Dark Matter Experiments}
\author{Marcela Carena$^{1}$, Dan Hooper$^{2}$, Alberto Vallinotto$^{2,3,4}$}
\affiliation{$^1$ Theoretical Physics, Fermi National Accelerator Laboratory, Batavia, IL  60510 \\
$^2$Particle Astrophysics Center, Fermi National Accelerator Laboratory, Batavia, IL  60510 \\
$^3$ Physics Department, The University of Chicago, Chicago, IL 60637 \\
$^4$ Institut d'Astrophysique de Paris, UMR7095 CNRS, Universite Pierre \& Marie Curie, 98 bis Boulevard Arago, 75014 Paris}
\date{\today}

\bigskip

\begin{abstract}
	
In this article, we explore the interplay between searches for supersymmetric particles and Higgs bosons at hadron colliders (the Tevatron and the LHC) and direct dark matter searches (such as CDMS, ZEPLIN, XENON, EDELWEISS, CRESST, WARP and others). We focus on collider searches for heavy MSSM Higgs bosons ($A$, $H$, $H^{\pm}$) and how the prospects for these searches are impacted by direct dark matter limits and vice versa. We find that the prospects of these two experimental programs are highly interrelated. A positive detection of $A$, $H$ or $H^{\pm}$ at the Tevatron would dramatically enhance the prospects for a near future direct discovery of neutralino dark matter. Similarly, a positive direct detection of neutralino dark matter would enhance the prospects of discovering heavy MSSM Higgs bosons at the Tevatron or the LHC. Combining the information obtained from both types of experimental searches will enable us to learn more about the nature of supersymmetry.
\end{abstract}

\pacs{PAC numbers: }
\maketitle

\section{Introduction}

Recently, direct dark matter detection experiments, most notably CDMS, have reached the level of sensitivity needed to detect neutralino dark matter over a substantial range of supersymmetric parameter space \cite{cdms,zeplin,edelweiss,cresst,warp}. The reach of these experiments continues to expand rapidly, probing an ever increasing fraction of supersymmetric models.

As searches for neutralino dark matter progress \cite{dmreview}, searches for supersymmetry at Run II of the Tevatron are becoming increasingly sensitive as greater integrated luminosity is accumulated. Results from Tevatron searches for neutralinos and charginos \cite{tevneucha}, MSSM Higgs bosons \cite{tevhiggs} and squarks and gluinos \cite{tevsquarksgluinos} have recently been published. Although no evidence for supersymmetry has yet been found at the Tevatron, many of these limits are the strongest to date and continue to advance. Furthermore, the first collisions at the Large Hadron Collider (LHC), with a center-of-mass energy of 14 TeV, are scheduled to take place in 2008. If low energy supersymmetry exists in nature, it will likely be discovered at these experiments in the relatively near future.

These two very different experimental programs - dark matter searches and collider experiments - are carried out largely independently of each other, with relatively little interaction between their respective communities. There is, however, a great deal of physics interplay between these two search strategies. Direct dark matter detection prospects depend on the mass and couplings of the lightest neutralino, as well as those of the Higgs bosons and squarks exchanged in elastic scattering diagrams. Many of these properties can potentially be measured or constrained by collider experiments. Turning this around, an astrophysical detection of neutralino dark matter, or an upper limit on its elastic scattering cross section with nucleons, can be used to provide information which is valuable to the accelerator community. 

In this paper, we expand on earlier work \cite{directsusypar}, which explored the interplay between direct dark matter experiments and searches for supersymmetric Higgs bosons at the Tevatron. In this article, we focus on Tevatron and LHC searches for heavy MSSM Higgs bosons ($A$, $H$, $H^{\pm}$), and discuss their relationship to direct neutralino dark matter searches. The paper is organized as follows. In Sec.~\ref{susy} we introduce the aspects of the supersymmetric spectrum relevant to our study. In Sec.~\ref{sec3} we review the contributions to the spin-independent neutralino-nucleon elastic scattering cross section and explore the impact on this cross section of the supersymmetric particle spectrum relevant to collider searches. In Sec.~\ref{sec4} we present the current exclusion limits and future discovery reach of the Tevatron and the LHC in the search for heavy MSSM neutral Higgs bosons, and compare this to the reach of direct neutralino dark matter searches. In Sec.~\ref{sec5} we discuss the implications and interplay of these two very different classes of experiments. In Sec.~\ref{sec6} we describe the caveats and limitations of our analysis. Finally, we summarize and state our conclusions in Sec.~\ref{sec7}.

\section{Neutralinos, Squarks and Higgs Bosons in the MSSM}
\label{susy}

In the Minimal Supersymmetric Standard Model (MSSM), the superpartners of the four Standard Model neutral bosons (the bino, wino and two neutral higgsinos) mix into four physical states known as neutralinos. The neutralino mass matrix in the $\widetilde{B}$-$\widetilde{W}$-$\widetilde{H}_1$-$\widetilde{H}_2$ basis is given by
\begin{equation}
\arraycolsep=0.01in
{\cal M}_N=\left( \begin{array}{cccc}
M_1 & 0 & -M_Z\cos \beta \sin \theta_W^{} & M_Z\sin \beta \sin \theta_W^{}
\\
0 & M_2 & M_Z\cos \beta \cos \theta_W^{} & -M_Z\sin \beta \cos \theta_W^{}
\\
-M_Z\cos \beta \sin \theta_W^{} & M_Z\cos \beta \cos \theta_W^{} & 0 & -\mu
\\
M_Z\sin \beta \sin \theta_W^{} & -M_Z\sin \beta \cos \theta_W^{} & -\mu & 0
\end{array} \right)\;,
\end{equation}
where $M_1$ and $M_2$ are the bino and wino masses, $\mu$ is the higgsino mass parameter, $\theta_W$ is the Weinberg angle, and $\tan \beta$ is the ratio of the vacuum expectation values of the up and down Higgs doublets. This matrix can be diagonalized into mass eigenstates by
\begin{equation}
M_{\chi^0}^{\rm{diag}} = N^{\dagger}  M_{\chi^0} N.
\end{equation}
We are interested here in the lightest neutralino, which in the presence of R-parity conservation can serve as a viable dark matter candidate.\footnote{R-parity is defined as $R=(-1)^{3B+L+2S}$, where $B$, $L$ and $S$ denote baryon number, lepton number and spin, respectively. $R=1$ for all SM particles and -1 for their superpartners.} In terms of the elements of the matrix, $N$, the lightest neutralino is given by the following mixture of gaugino and higgsino components:
\begin{equation}
\chi^0 = N_{11}\tilde{B}     +N_{12} \tilde{W}^3
          +N_{13}\tilde{H}_1 +N_{14} \tilde{H}_2.
\label{eq3}
\end{equation}
The mass and composition of the lightest neutralino is a function of four supersymmetric parameters: $M_1$, $M_2$, $\mu$ and $\tan \beta$. This becomes further simplified if the gaugino masses are assumed to evolve to a single value at the GUT scale, yielding a ratio at the electroweak scale of $M_1 = \frac{5}{3} \tan^2\theta_W M_2 \approx 0.5 M_2 $.

As we will describe in the following section, neutralinos can scatter with nuclei through the s-channel exchange of the superpartners of the Standard Model quarks. These contribution can be particularly important in the case of light squark masses and large to moderate values of $\tan \beta$. The diagonal entries in the squark mass matrices are largely set by the soft supersymmetry breaking mass parameters $m_{Q_i}$, $m_{U_i}$ and $m_{D_i}$, where $i=1,2,3$ denotes the generation. In many supersymmetry breaking scenarios, such as minimal supergravity and gauge mediated models, these parameters are nearly degenerate, leading to all of the first and second generation squarks to naturally have very similar masses. Third generation squarks (stops and sbottoms) have their masses split by their Yukawa couplings, however. The stop and sbottom mass matrices are given by:

\begin{equation}
\arraycolsep=0.01in
{\cal M}^2_{\tilde{t}}=\left( \begin{array}{cc}
m^2_{Q_3}+ m^2_t + D_{\tilde{u}_L}\,\, &\,\, m_t \, (A_t-\mu \cot\beta) \\
m_t \, (A_t-\mu \cot\beta)\,\, & \,\,m^2_{U_3}+ m^2_t + D_{\tilde{u}_R}\\
\end{array} \right)\;
\end{equation}
and
\begin{equation}
\arraycolsep=0.01in
{\cal M}^2_{\tilde{b}}=\left( \begin{array}{cc}
m^2_{Q_3}+ m^2_b + D_{\tilde{d}_L}\,\, &\,\, m_b \, (A_b-\mu \tan\beta) \\
m_b \, (A_b-\mu \tan\beta)\,\, & \,\,m^2_{D_3}+ m^2_b + D_{\tilde{d}_R}\\
\end{array} \right)
\end{equation}
Here, the $D$'s are D-term contributions, and are of order $m^2_Z$. The quantities $A_t$ and $A_b$ are soft supersymmetry breaking trilinear scalar couplings in one-to-one correspondence to the Yukawa terms in the superpotential. The mass splitting in the stop sector can be especially pronounced. Defining a common squark mass $m_{\tilde{q}} \equiv m_{Q_i}=m_{U_i}=m_{D_i}$, then for $m_{\tilde{q}} \gg m_t$, the physical stop masses are well approximated by $m_{\tilde{t}_{1,2}} \approx [m^2_{\tilde{q}} \mp m_t (A_t-\mu \cot \beta) ]^{1/2}$.

Throughout this study, we will consider two benchmark scenarios which provide conventional choices for the mixing in the stop sector~\cite{maxnomixing}. The first of these, called the ``no-mixing scenario'', is defined by $A_t=\mu \cot \beta$, minimizing the mixing in the stop sector and yielding the minimal radiative corrections to the lightest CP-even Higgs mass, $m_h$. The second case, called the ``$m^{\rm{max}}_h$ scenario'', is defined by $A_t=2 m_{\tilde{q}} + \mu \cot \beta$, and maximizes the value of $m_h$. For definiteness, we adopt $A_b=A_t$ throughout.

In addition to exchanging squarks, the elastic scattering of neutralinos with nuclei can be induced through the exchange of Higgs bosons. Rather than the single Higgs boson of the Standard Model, a minimum of two Higgs doublets are required in supersymmetric models to avoid triangle diagram anomalies and to create a gauge-invariant superpotential that gives masses to both up and down-type quark fermions. In the absence of explicit CP-violation, the two Higgs doublets of the MSSM correspond to physical states in the form of two neutral, CP-even Higgs bosons ($h$ and $H$), one neutral CP-odd Higgs boson ($A$) and one charged Higgs boson ($H^{\pm}$). The angle $\alpha$ diagonalizes the CP-even Higgs squared mass matrix.
 
It is traditional to take $m_A$ as a free parameter which, together with the value of $\tan \beta$, determines at tree level the masses of all the other Higgs bosons:
\begin{eqnarray}
m^2_{H^{\pm}} &=& m^2_A+m^2_{W} \\
m^2_{h, H} &=& \frac{1}{2} \bigg(m^2_A + m^2_Z \mp \sqrt{(m^2_A + m^2_Z)^2 - 4 m_Z^2 m_A^2 \cos^2 2 \beta} \bigg)
\end{eqnarray}
In the case that $m_A \gg m_Z$, it follows that $m_H \approx m_{H^{\pm}} \approx m_A$ and $\cos \alpha \approx 1$. These simple relationships are significantly modified at the loop level, however, and become strongly dependent on the supersymmetric parameters in the stop and sbottom sectors. In our study, we have used the Feynhiggs package \cite{feynhiggs} to calculate the Higgs masses.\footnote{Similar results can be obtained using the CPsuperH package \cite{cpsuperh}.} For a more complete discussion of Higgs and supersymmetry phenomenology, see Refs.~\cite{primer}, \cite{susyreview2} and \cite{higgsreview}.

\vspace{1.0cm}

\section{Direct Detection of Neutralino Dark Matter}
\label{sec3}

Experiments such as CDMS \cite{cdms}, ZEPLIN \cite{zeplin}, EDELWEISS \cite{edelweiss}, CRESST \cite{cresst} and WARP \cite{warp} attempt to detect dark matter particles through their elastic scattering with nuclei. This class of techniques are collectively referred to as direct detection, in contrast to indirect detection efforts which attempt to observe the annihilation products of dark matter particles.

Neutralinos can scatter with nuclei through both scalar (spin-independent) and axial-vector (spin-dependent) interactions. The experimental sensitivity to scalar couplings benefits from coherent scattering, which leads to cross sections and rates proportional to the square of the atomic mass of the target nuclei. The cross sections for axial-vector elastic scattering are proportional to $J(J+1)$, however, and thus do not benfit from large target nuclei. As a result, the current experimental sensitivity to axial-vector couplings is far below that of scalar interactions, and well below the range predicted for neutralinos. For this reason, we consider only the scalar case here. 

The scalar neutralino-nuclei elastic scattering cross section is given by:
\begin{equation}
\label{sig}
\sigma \approx \frac{4 m^2_{\chi^0} m^2_{T}}{\pi (m_{\chi^0}+m_T)^2} [Z f_p + (A-Z) f_n]^2,
\end{equation}
where $m_T$ is the target nuclei's mass, and $Z$ and $A$ are the atomic number and atomic mass of the nucleus.  $f_p$ and $f_n$ are the neutralino couplings to protons and neutrons, given by:
\begin{equation}
f_{p,n}=\sum_{q=u,d,s} f^{(p,n)}_{T_q} a_q \frac{m_{p,n}}{m_q} + \frac{2}{27} f^{(p,n)}_{TG} \sum_{q=c,b,t} a_q  \frac{m_{p,n}}{m_q},
\label{feqn}
\end{equation}
where $a_q$ are the neutralino-quark couplings and, being conservative, $f^{(p)}_{T_u} \approx 0.020\pm0.004$,  
$f^{(p)}_{T_d} \approx 0.026\pm0.005$,  $f^{(p)}_{T_s} \approx 0.118\pm0.062$,  
$f^{(n)}_{T_u} \approx 0.014\pm0.003$,  $f^{(n)}_{T_d} \approx 0.036\pm0.008$ and 
$f^{(n)}_{T_s} \approx 0.118\pm0.062$ \cite{nuc}.

The first term in Eq.~\ref{feqn} 
corresponds to interactions with the quarks in the target nuclei, either through $t$-channel CP-even Higgs exchange, or $s$-channel squark exchange:
\vspace{-0.2cm}

\begin{feynartspicture}(222,254)(3,4.3)
\FADiagram{ }
\FAProp(20.,20.)(35.,10.)(0.,){/Straight}{0}
\FALabel(15.,18.0)[b]{$\chi^0$}
\FAProp(35.,-5.0)(35.,10.0)(0.,){/ScalarDash}{0}
\FAProp( 20.,-15.0)(35.,-5.0)(0.,){/Straight}{+1}
\FALabel(15.,-18.0)[b]{$q$}
\FALabel(40.,0.0)[b]{$H, h$}
\FAProp(35.,10.)(50.,20.)(0.,){/Straight}{0}
\FAProp(35.,-5.)(50.,-15.0)(0.,){/Straight}{+1}
\FALabel(55.,18.0)[b]{$\chi^0$}
\FALabel(55.,-18.)[b]{$q$}



\FAProp(90.,14.)(105.,3.)(0.,){/Straight}{0}
\FALabel(85.,12.0)[b]{$\chi^0$}

\FAProp(90.,-7.0)(105.,3.0)(0.,){/Straight}{+1}
\FALabel(85.,-9.0)[b]{$q$}

\FAProp(105.,3.)(125.,3.0)(0.,){/ScalarDash}{0}
\FALabel(115.,9.)[t]{$\tilde{q}$}

\FAProp(140.,14.)(125.,3.0)(0.,){/Straight}{0}
\FALabel(145.,12.)[b]{$\chi^0$}

\FAProp(140.,-7.)(125.,3.0)(0.,){/Straight}{+1}
\FALabel(145.,-9.)[b]{$q$}
\end{feynartspicture}

\vspace{-4.5cm}
\noindent
The second term corresponds to interactions with the gluons in the target through a quark/squark 
loop diagram. $f^{(p)}_{TG}$ is given by $1 -f^{(p)}_{T_u}-f^{(p)}_{T_d}-f^{(p)}_{T_s} 
\approx 0.84$, and analogously, $f^{(n)}_{TG} \approx 0.83$. To account for finite momentum transfer, the calculation should also include the appropriate form factor.

The neutralino-quark coupling, in which all of the SUSY model-dependent information is contained, is given by~\cite{scatteraq}:
\begin{eqnarray}
\label{aq}
a_q & = & - \frac{1}{2(m^{2}_{1i} - m^{2}_{\chi})} Re \left[
\left( X_{i} \right) \left( Y_{i} \right)^{\ast} \right] 
- \frac{1}{2(m^{2}_{2i} - m^{2}_{\chi})} Re \left[ 
\left( W_{i} \right) \left( V_{i} \right)^{\ast} \right] \nonumber \\
& & \mbox{} - \frac{g_2 m_{q}}{4 m_{W} B} \left[ Re \left( 
\delta_{1} [g_2 N_{12} - g_1 N_{11}] \right) D C \left( - \frac{1}{m^{2}_{H}} + 
\frac{1}{m^{2}_{h}} \right) \right. \nonumber \\
& & \mbox{} +  Re \left. \left( \delta_{2} [g_2 N_{12} - g_1 N_{11}] \right) \left( 
\frac{D^{2}}{m^{2}_{h}}+ \frac{C^{2}}{m^{2}_{H}} 
\right) \right],
\end{eqnarray}
where
\begin{eqnarray}
X_{i}& \equiv& \eta^{\ast}_{11} 
        \frac{g_2 m_{q}N_{1, 5-i}^{\ast}}{2 m_{W} B} - 
        \eta_{12}^{\ast} e_{i} g_1 N_{11}^{\ast}, \nonumber \\
Y_{i}& \equiv& \eta^{\ast}_{11} \left( \frac{y_{i}}{2} g_1 N_{11} + 
        g_2 T_{3i} N_{12} \right) + \eta^{\ast}_{12} 
        \frac{g_2 m_{q} N_{1, 5-i}}{2 m_{W} B}, \nonumber \\
W_{i}& \equiv& \eta_{21}^{\ast}
        \frac{g_2 m_{q}N_{1, 5-i}^{\ast}}{2 m_{W} B} -
        \eta_{22}^{\ast} e_{i} g_1 N_{11}^{\ast}, \nonumber \\
V_{i}& \equiv& \eta_{22}^{\ast} \frac{g_2 m_{q} N_{1, 5-i}}{2 m_{W} B}
        + \eta_{21}^{\ast}\left( \frac{y_{i}}{2} g_1 N_{11},
        + g_2 T_{3i} N_{12} \right),
\label{xywz}
\end{eqnarray}
where throughout $i=1$ for up-type quarks and $i=2$ for down type quarks. $m_{1i}, m_{2i}$ denote the squark mass eigenvalues and $\eta$ is the matrix which diagonalizes the squark mass matrices, $diag(m^2_1, m^2_2)=\eta M^2 \eta^{-1}$. $y_i$, $T_{3i}$ and $e_i$ denote hypercharge, isospin and electric charge of the quarks. For scattering off of up-type quarks:
\begin{eqnarray}
\delta_{1} = N_{13},\,\,\,\, \delta_{2} = N_{14}, \,\,\,\, B = \sin{\beta},\,\,\,\, C = \sin{\alpha}, \,\,\,\, D = \cos{\alpha},
\end{eqnarray}
whereas for down-type quarks:
\begin{eqnarray}
\delta_{1} = N_{14},\,\,\,\, \delta_{2} = -N_{13}, \,\,\,\, B = \cos{\beta},\,\,\,\, C = \cos{\alpha}, \,\,\,\, D = -\sin{\alpha}.
\end{eqnarray}

The first two terms of Eq.~\ref{aq} correspond to interactions through the exchange of a squark, while the final term is generated through Higgs exchange.

We will now describe the behavior of this cross section in a number of interesting limiting cases. Firstly, in the case of heavy squarks, small wino component ($N_{12}$) and $\cos \alpha \approx 1$ (which implies moderate to large $\tan \beta$ and $m_A \sim m_H \sim m_{H^{\pm}}$), elastic scattering is dominanted by $H$ exchange with strange and bottom quarks, leading to a neutralino-nucleon cross section of:
\begin{equation}
\sigma_{\chi N} \sim \frac{g^2_1 g^2_2 |N_{11}|^2 |N_{13}|^2 \,m^4_N}{4\pi m^2_W \cos^2 \beta \, m^4_H} \bigg(f_{T_s}+\frac{2}{27}f_{TG}\bigg)^2, \,\,\,\, (m_{\tilde{q}}\, \rm{large}, \cos \alpha \approx 1).
\label{case1}
\end{equation}

If instead, we consider the case of heavy squarks, small $N_{12}$ and heavy $H$ ($\cos \alpha \approx 1$), we find that scattering with up-type quarks through light Higgs exchange dominates:
\begin{equation}
\sigma_{\chi N} \sim \frac{g^2_1 g^2_2 |N_{11}|^2 |N_{14}|^2 \, m^4_N}{4\pi m^2_W \, m^4_h} \bigg(f_{T_u}+\frac{4}{27}f_{TG}\bigg)^2, \,\,\,\, (m_{\tilde{q}}, m_H\, \rm{large}, \cos \alpha \approx 1).
\label{case2}
\end{equation}

Thirdly, for the case of scattering dominanted by squark exchange, with large to moderate $\tan \beta$ and approximately diagonal squark mass matrices, we find the limit of:
\begin{equation}
\sigma_{\chi N} \sim \frac{g^2_1 g^2_2 |N_{11}|^2 |N_{13}|^2 \, m^4_N}{4\pi m^2_W \cos^2 \beta \, m^4_{\tilde{q}}} \bigg(f_{T_s}+\frac{2}{27}f_{TG}\bigg)^2, \,\,\,\, (\tilde{q}\,\, \rm{dominated}, \tan \beta \gg 1).
\label{case3}
\end{equation}

\begin{figure}[t]
\hspace{-1.0cm}
\includegraphics[width=3.65in,angle=0]{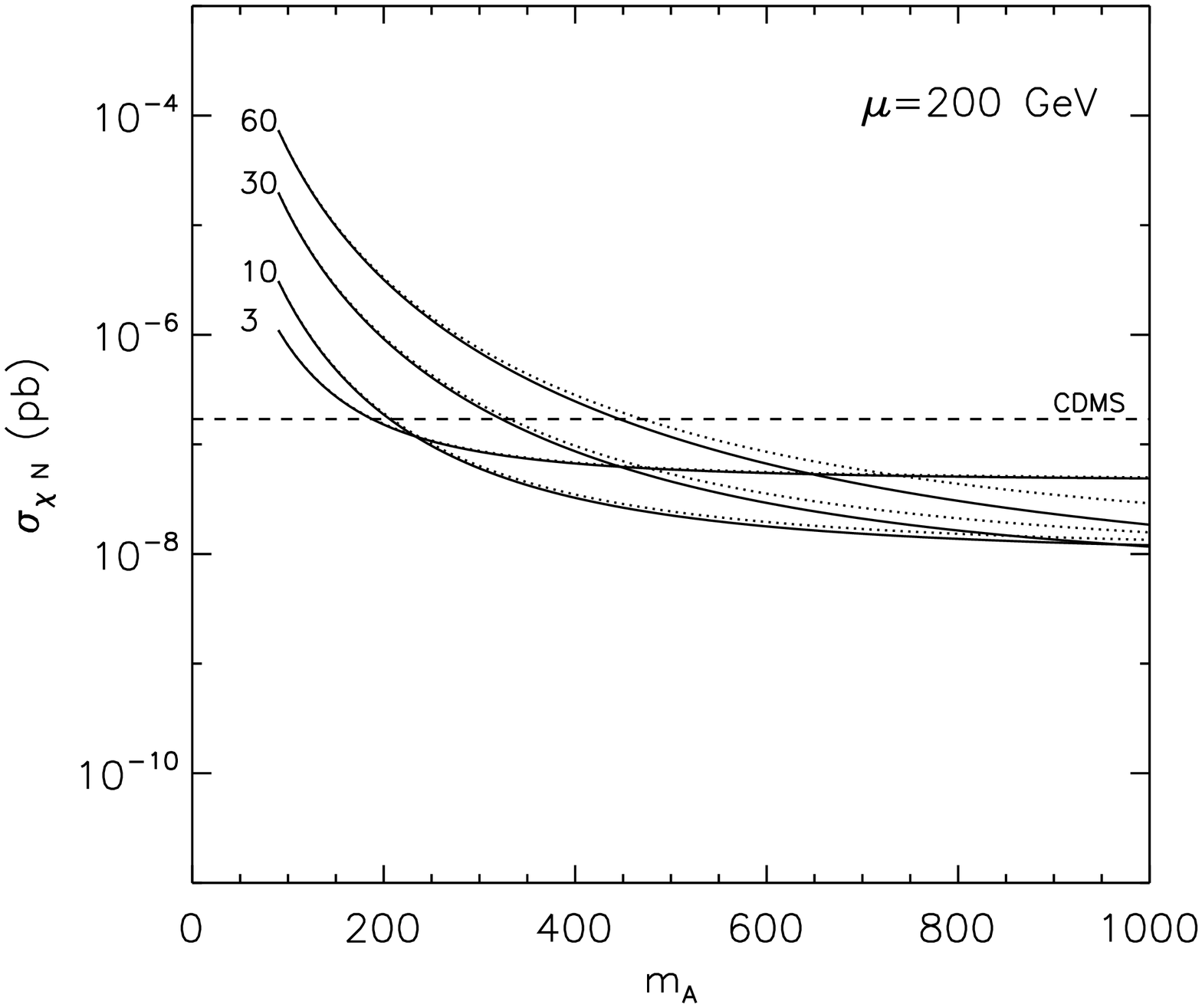}
\hspace{-1.1cm}
\includegraphics[width=3.65in,angle=0]{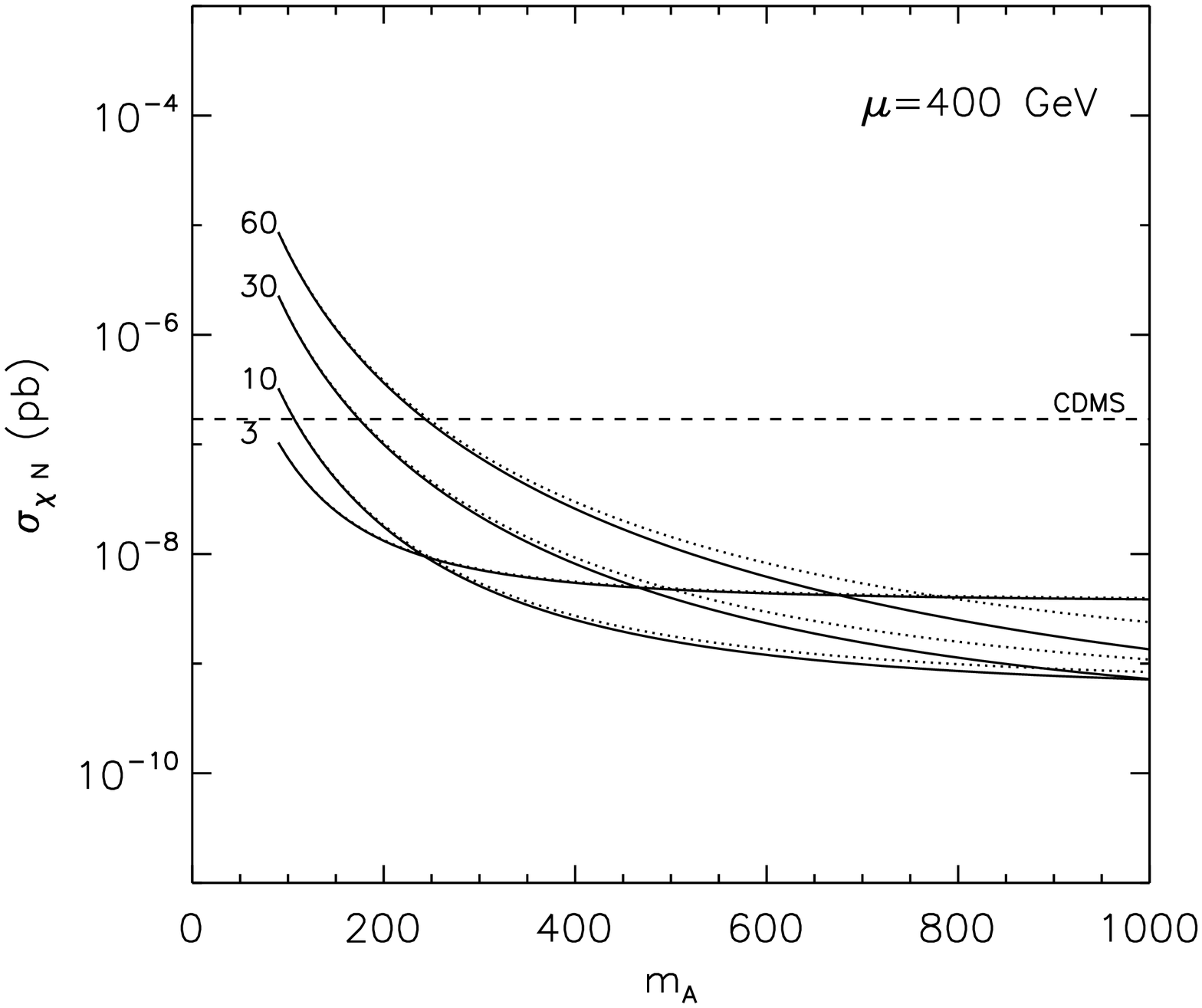}\\
\hspace{-1.0cm}
\includegraphics[width=3.65in,angle=0]{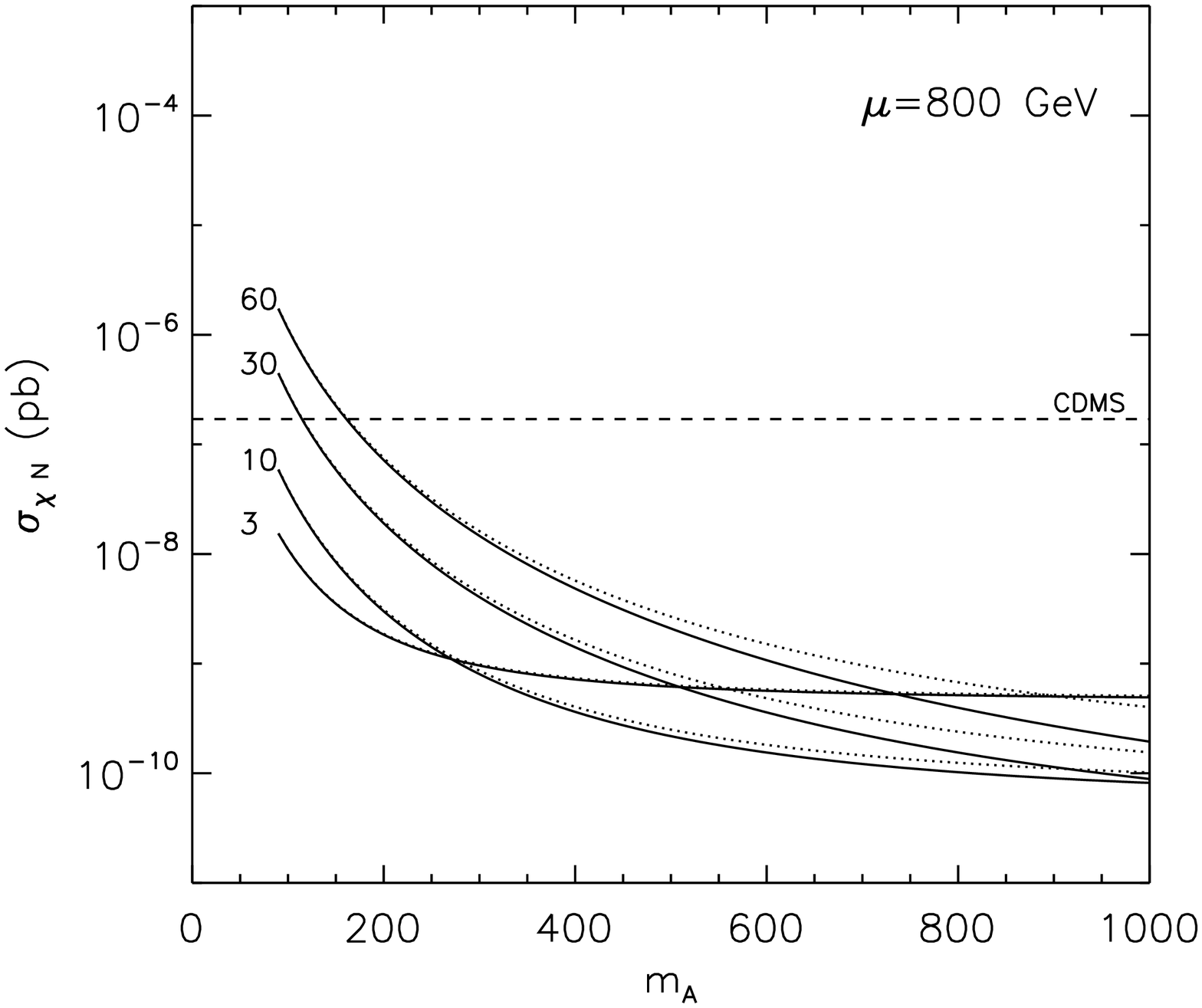}
\hspace{-1.1cm}
\includegraphics[width=3.65in,angle=0]{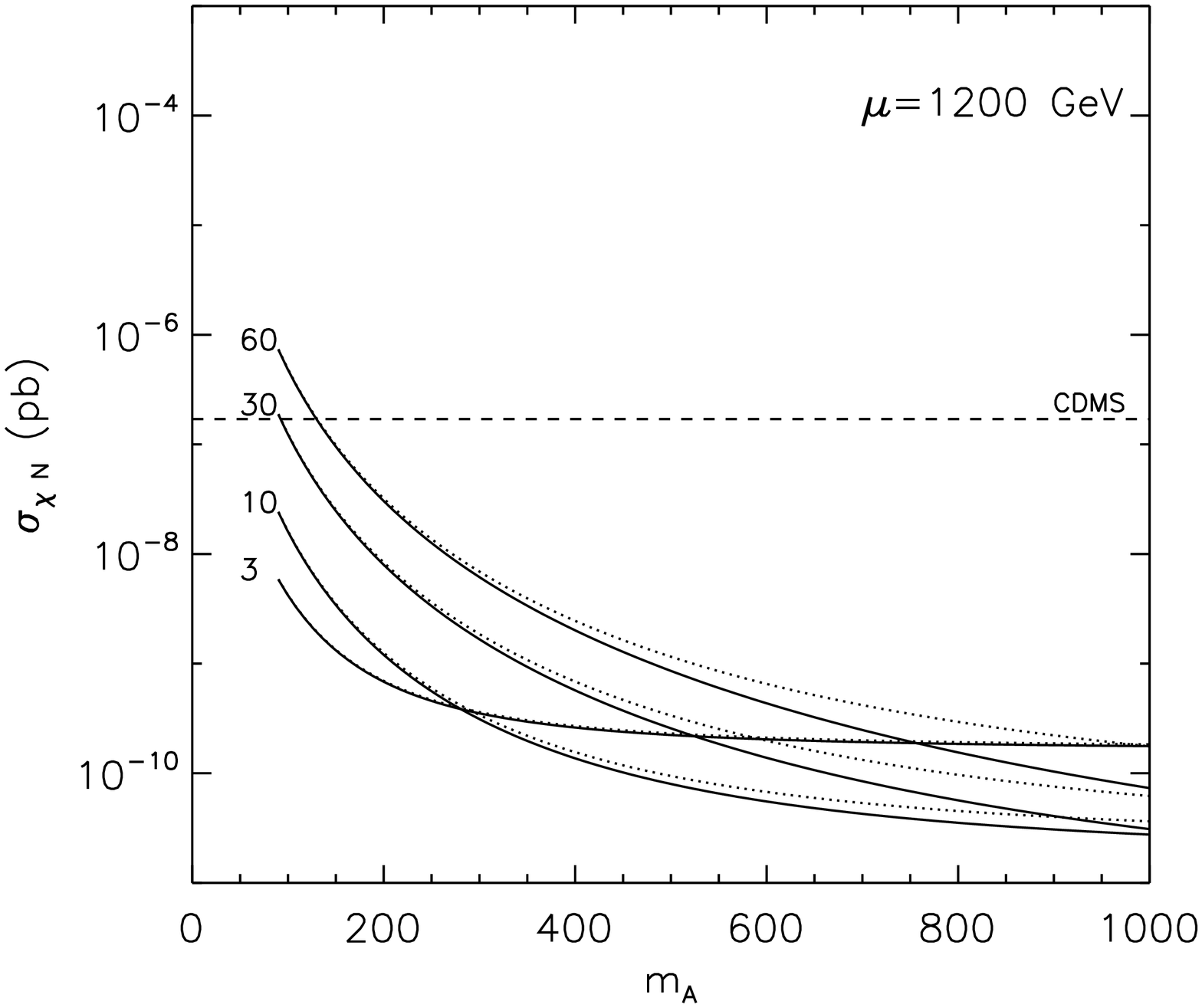}
\caption{The neutralino-proton spin-independent elastic scattering cross section as a function of $m_A$ for selected values of $\mu$ (200, 400, 800 and 1200 GeV) and $\tan \beta$ (3, 10, 30 and 60). For each case, $M_1=100$ GeV, $M_2=200$ GeV, $A_t=A_b=\mu \cot\beta$ (the no-mixing scenario) and $m_{\tilde{q}}=1$ TeV have been used. The results found for $A_t=A_b= 2 m_{\tilde{q}} +\mu \cot \beta$ (the $m^{\rm{max}}_{h}$ scenario) are very similar. The solid lines {\it do not} include any contributions from squark exchange, while the dotted lines include both Higgs and squark exchange. Shown as a horizontal dashed line is the current upper limit from the CDMS experiment~\cite{cdms}.}
\label{sigmahiggs1}
\end{figure}

\begin{figure}[!t]
\hspace{-1.0cm}
\includegraphics[width=3.65in,angle=0]{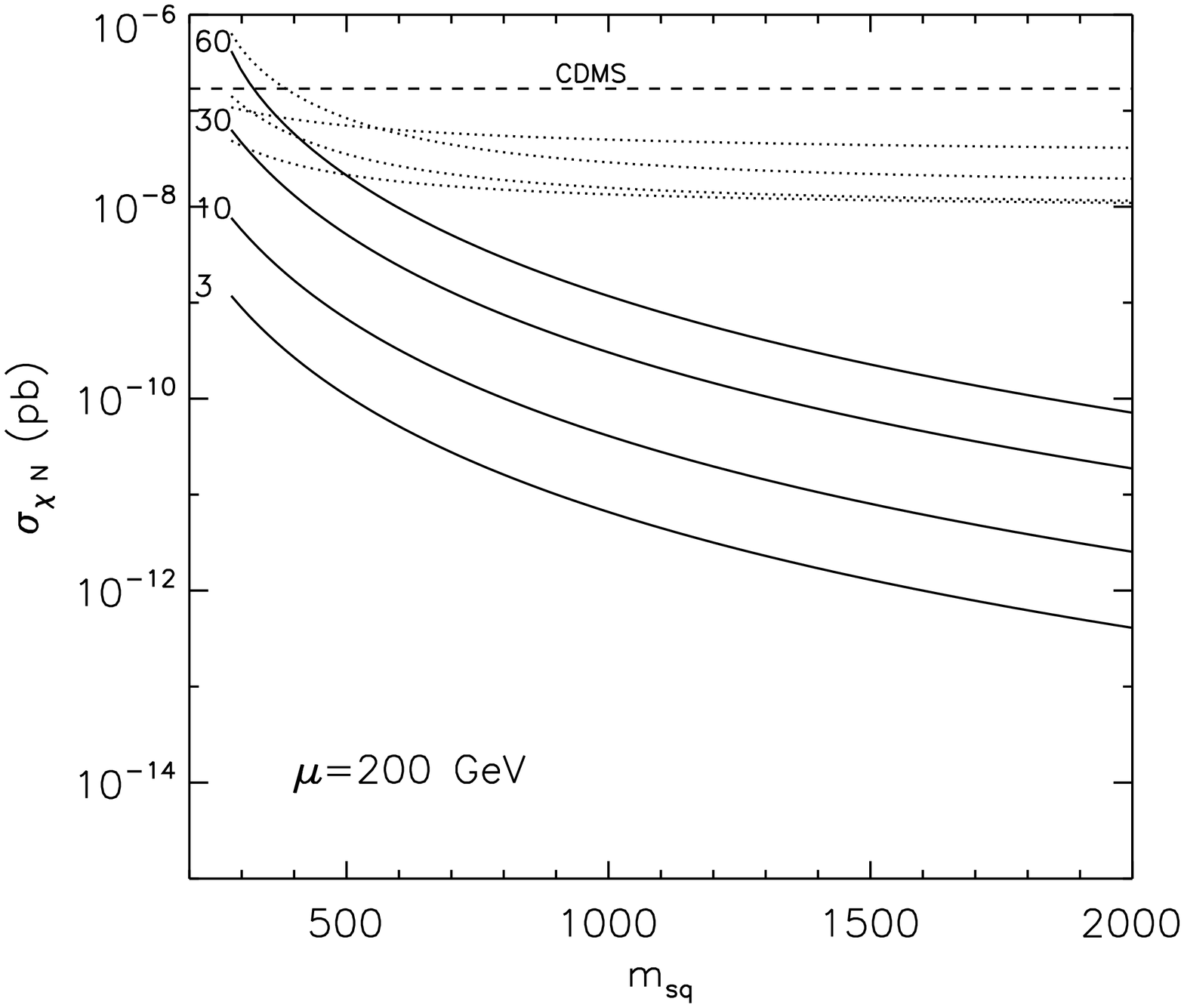}
\hspace{-1.1cm}
\includegraphics[width=3.65in,angle=0]{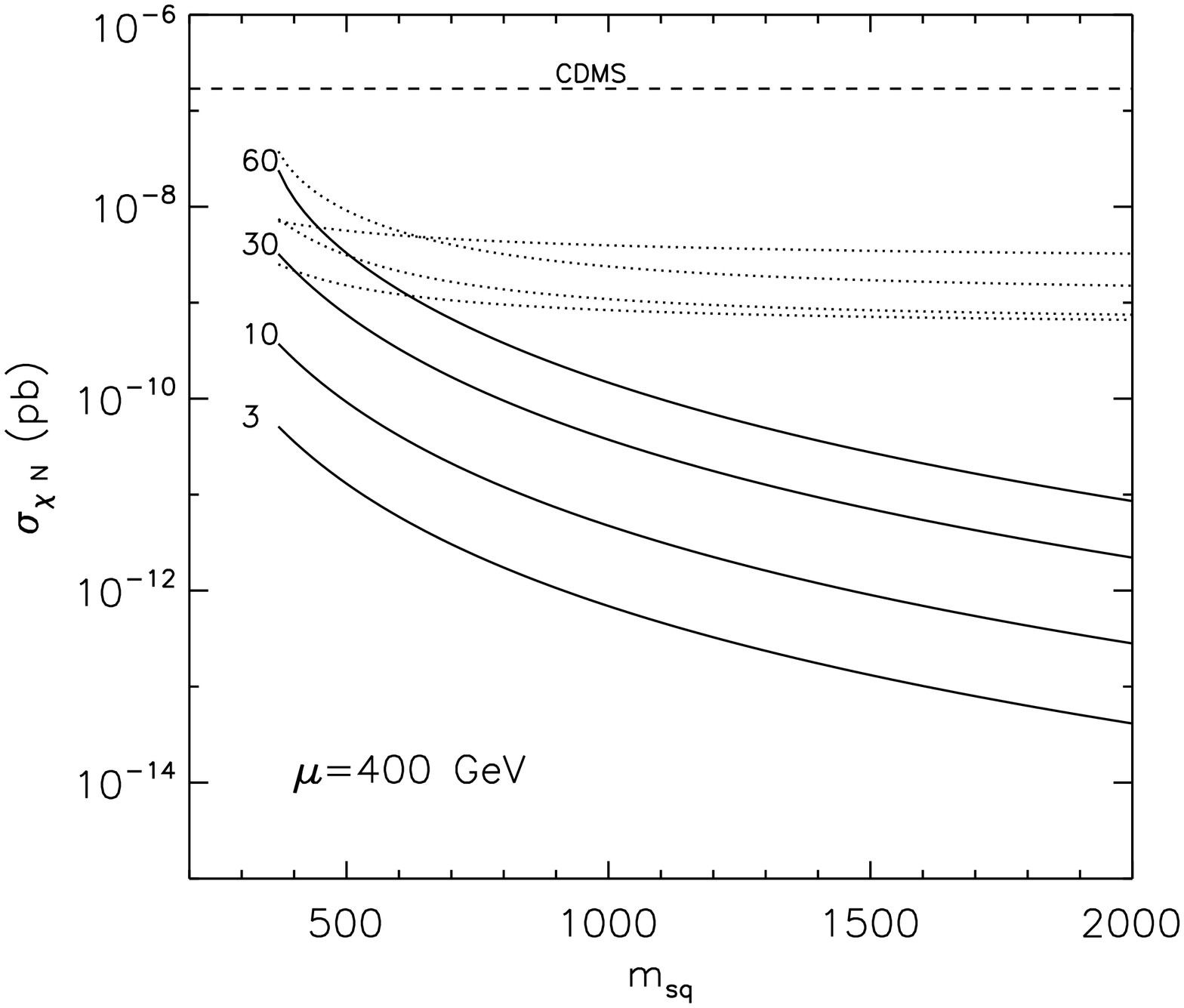}\\
\hspace{-1.0cm}
\includegraphics[width=3.65in,angle=0]{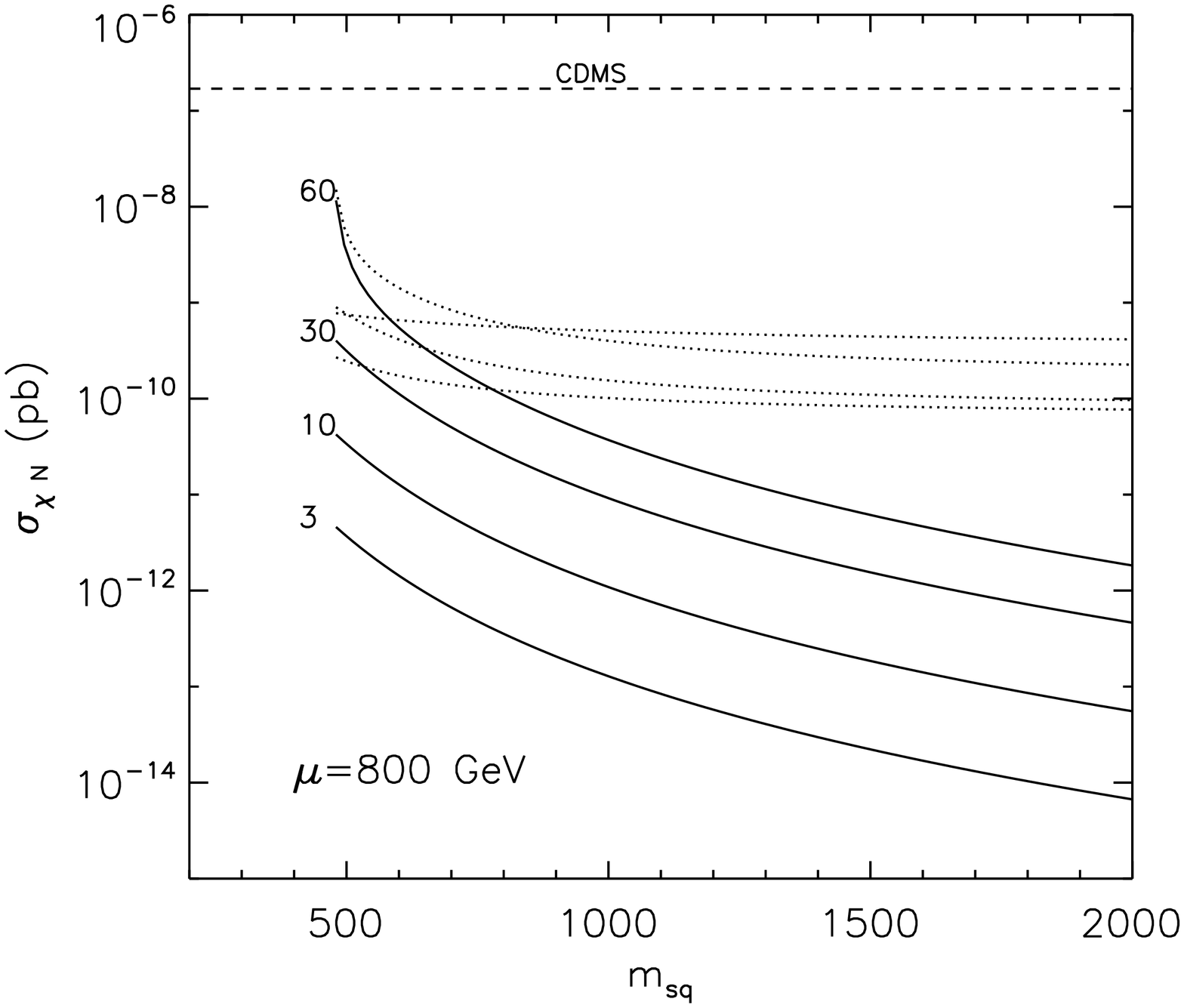}
\hspace{-1.1cm}
\includegraphics[width=3.65in,angle=0]{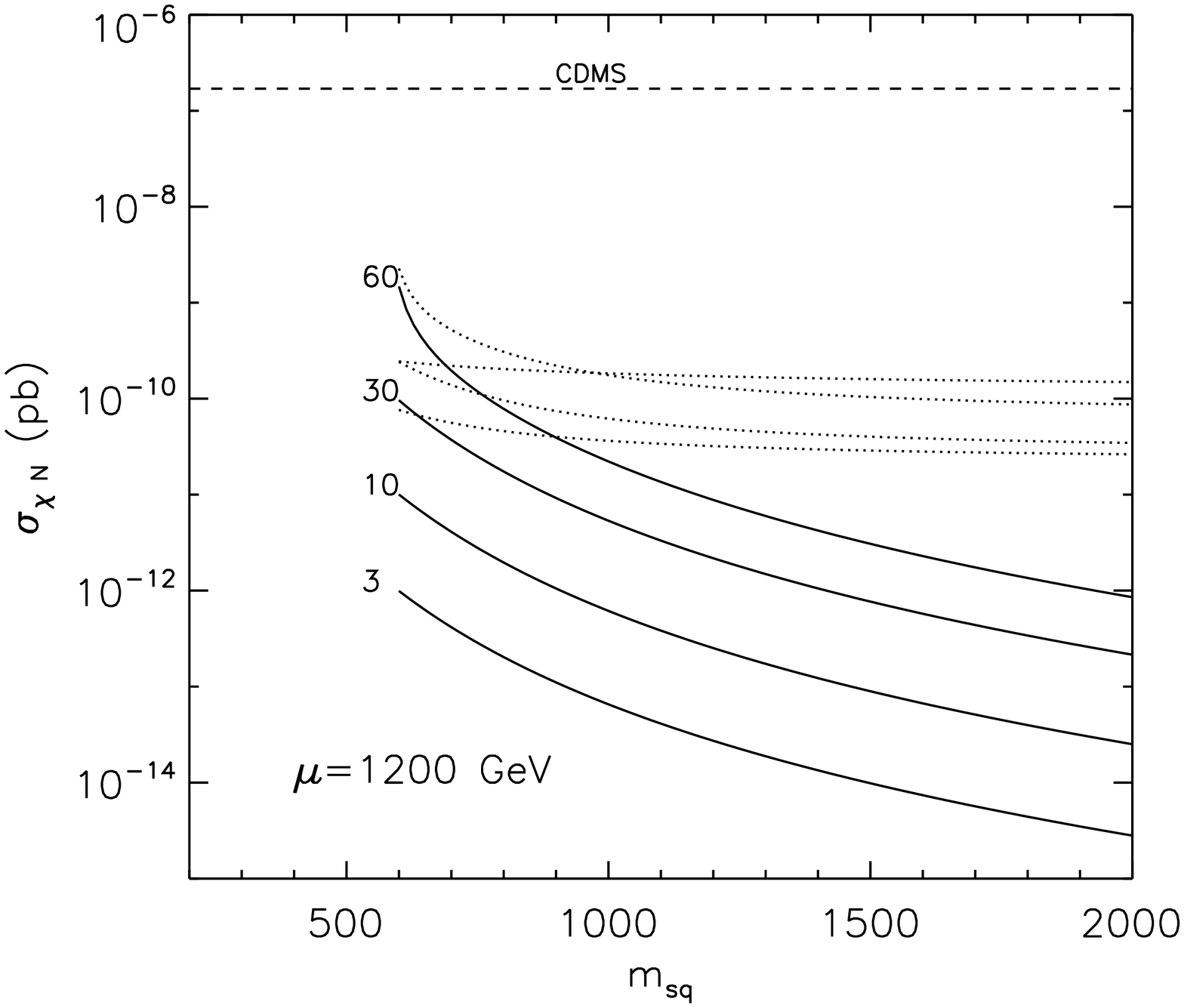}
\caption{The neutralino-proton spin-independent elastic scattering cross section as a function of $m_{\tilde{q}}$ for selected values of $\mu$ (200, 400, 800 and 1200 GeV) and $\tan \beta$ (3, 10, 30 and 60). For each case, $M_1=100$ GeV, $M_2=200$ GeV,  $A_t=A_b= \mu \cot \beta$ (the no-mixing scenario) and $m_A=1$ TeV have been used.  The results found for $A_t=A_b= 2 m_{\tilde{q}} +\mu \cot \beta$ (the $m^{\rm{max}}_{h}$ scenario) are very similar. The solid lines {\it do not} include any contributions from Higgs exchange, while the dotted lines include both Higgs and squark exchange. Shown as a horizontal dashed line is the current upper limit from the CDMS experiment~\cite{cdms}.}
\label{sigmasquark1}
\end{figure}

We can see these behaviors exhibited in Fig.~\ref{sigmahiggs1} where we plot the neutralino-nucleon spin-independent elastic scattering cross section as a function of the CP-odd Higgs mass, $m_A$ (recall, that for $m_A \gg m_Z$, $m_A \approx m_H$). In this figure we have adopted the no-mixing scenario and consider different values of $\tan \beta$ and $\mu$.\footnote{Note that in the case of the no-mixing scenario, the $\tan \beta=3$ contours appearing in Figs.~\ref{sigmahiggs1} and \ref{sigmasquark1} predict a value of $m_h$ which is excluded by LEP.} For relatively small values of $m_A$, the cross section scales with $1/m^4_A$, as found in the limit of Eq.~\ref{case1}. As $m_A$ becomes larger, the cross section becomes dominanted by $h$ exchange and flattens with respect to $m_A$, as found in the limit of Eq.~\ref{case2}. 

For large $\tan \beta$, large $m_A$ and light squark masses (the limit of Eq.~\ref{case3}), the effect of the squark exchange contribution on the elastic scattering cross section becomes significant. In Fig.~\ref{sigmasquark1} we plot the neutralino-nucleon spin-independent elastic scattering cross section for various values of $\tan \beta$ and $\mu$ as a function of the squark masses. From this we see that in the case that the $H$ exchange contribution is suppressed (due to large $m_A \approx m_H$), squark exchange diagrams dominate the elastic scattering cross section if the squarks are light and $\tan \beta$ is large.

\clearpage	
\section{Heavy MSSM Higgs Boson Searches at the Tevatron and LHC}
\label{sec4}

Searches for heavy MSSM Higgs bosons are being carried out at Run II of the Tevatron. In particular, efforts are underway to observe the processes:
\begin{eqnarray}
&p\bar{p}& \rightarrow A/H + X \rightarrow \tau^+ \tau^- +X, \nonumber \\ 
&p\bar{p}& \rightarrow A/H + b \bar{b} \rightarrow  b \bar{b} + b \bar{b},\,\, \rm{(3 }b\rm{'s}\,\,\rm{tagged)} \nonumber \\
&p\bar{p}& \rightarrow t\bar{t} \rightarrow H^{\pm} + W^{\mp}\, b \bar{b} \rightarrow \tau^{\pm} \nu_{\tau} +  W^{\mp}\, b \bar{b}, \nonumber \\
&p\bar{p}& \rightarrow A/H + b \rightarrow \tau^+ \tau^- + b.
\label{tevhiggs}
\end{eqnarray}
At the LHC, such particles may be observed via
\begin{eqnarray}
&pp& \rightarrow A/H + X \rightarrow \tau^+ \tau^- +X, \\
&pp& \rightarrow H^{\pm} + t\,X \rightarrow \tau^{\pm} \nu_{\tau} + t\, X.
\end{eqnarray}

In each of these cases, the prospects for discovery are much greater in the case of large $\tan \beta$ and relatively light $m_A$. In the large $\tan \beta$ regime, the leading contributions to the production of an $A$ or $H$ rely on the b-quark Yukawa coupling, and thus scale with $\tan^2 \beta$. For example, $A/H$ production via gluon fusion is dominanted by diagrams with a b-quark loop. The cross section for charged Higgs production at the LHC also scales with $\tan^2 \beta$.

The current limits on heavy MSSM Higgs bosons from Run II of the Tevatron have been published for each of the channels shown in Eq.~\ref{tevhiggs}. CDF has published limits for the di-tau and charged Higgs channels using their first 310 and 193 pb$^{-1}$, respectively~\cite{higgscdfcurrent}. D0 has published (or presented) limits on the inclusive di-tau, $b\bar{b}$+$b\bar{b}$ and $b$-tau-tau channels using their first 348, 260 and 344 pb$^{-1}$, respectively \cite{higgsdzerocurrent}. The Tevatron currently has approximately 1 fb$^{-1}$ of data, and is expected to accumulate a total of 4 fb$^{-1}$ or more by the end of its operation.

The results of the heavy MSSM Higgs searches are generally represented in the $\tan \beta$--$m_A$ plane, for a given choice of $\mu$ and other SUSY parameters (which impact the limit through radiative corrections \cite{carena}). The limits found for the di-tau channel, at both the Tevatron and the LHC, are the most interesting and are quite robust to variations in $\mu$ and other supersymmetric parameters. In the special case of negative and large $\mu$, the limits from the b-quark channel at the Tevatron can be somewhat more constraining than the di-tau channel~\cite{carena}, although this is disfavored by the combination of measurements of the muon's magnetic moment \cite{gminus2,gminus2theory} and the $B \rightarrow X_s \gamma$ branching fraction \cite{bsg,bsgtheory}. The limits on charged Higgs bosons from the Tevatron, and as projected from the LHC, yield weaker constraints in the $\tan \beta$-$m_A$ plane at this time. For these reasons, we here focus on the di-tau channel for heavy Higgs searches at both colliders. 

\begin{figure}[!t]
\hspace{-1.0cm}
\includegraphics[width=3.25in,angle=0]{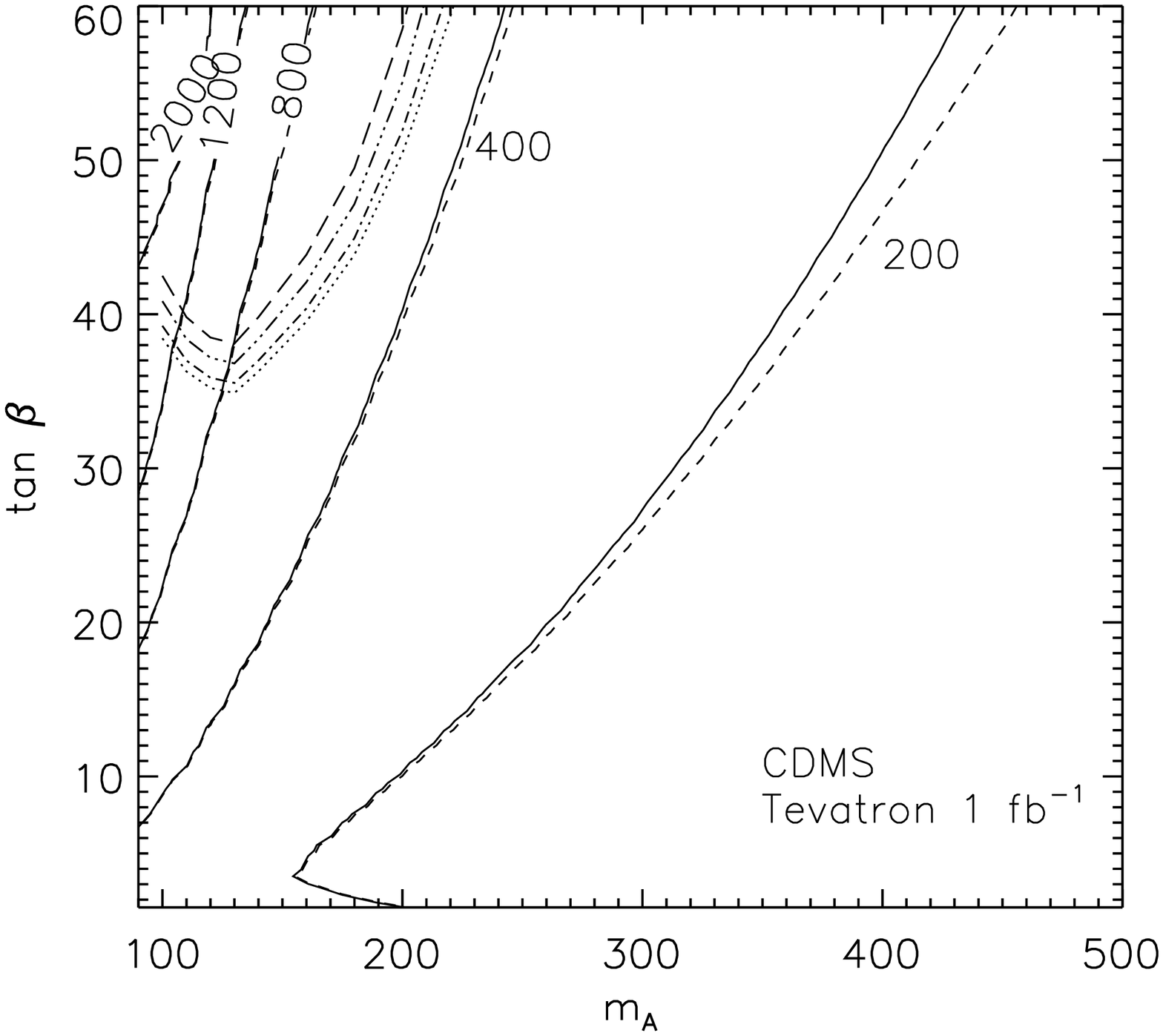}
\hspace{-1.1cm}
\includegraphics[width=3.25in,angle=0]{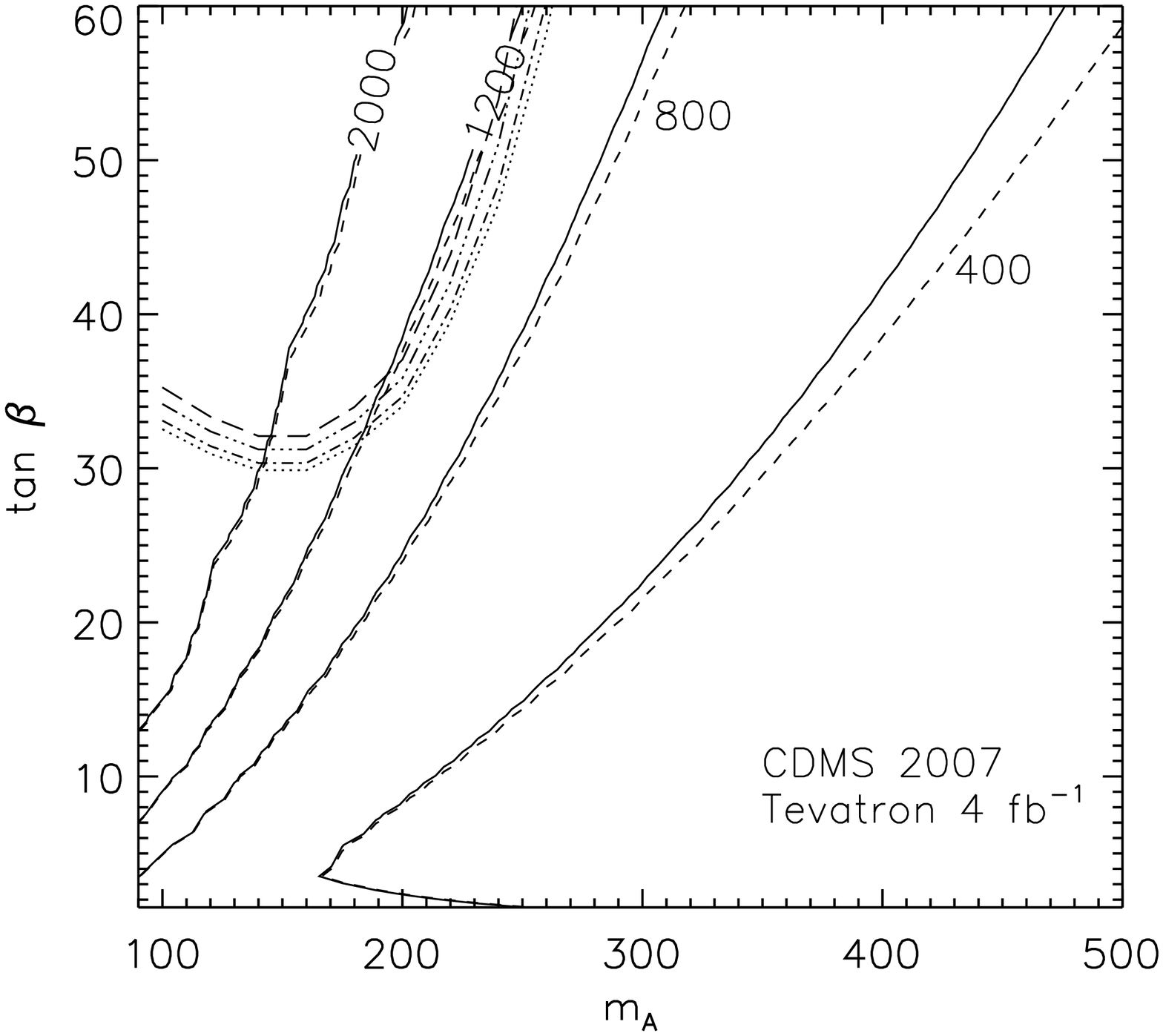}\\
\hspace{-1.0cm}
\includegraphics[width=3.25in,angle=0]{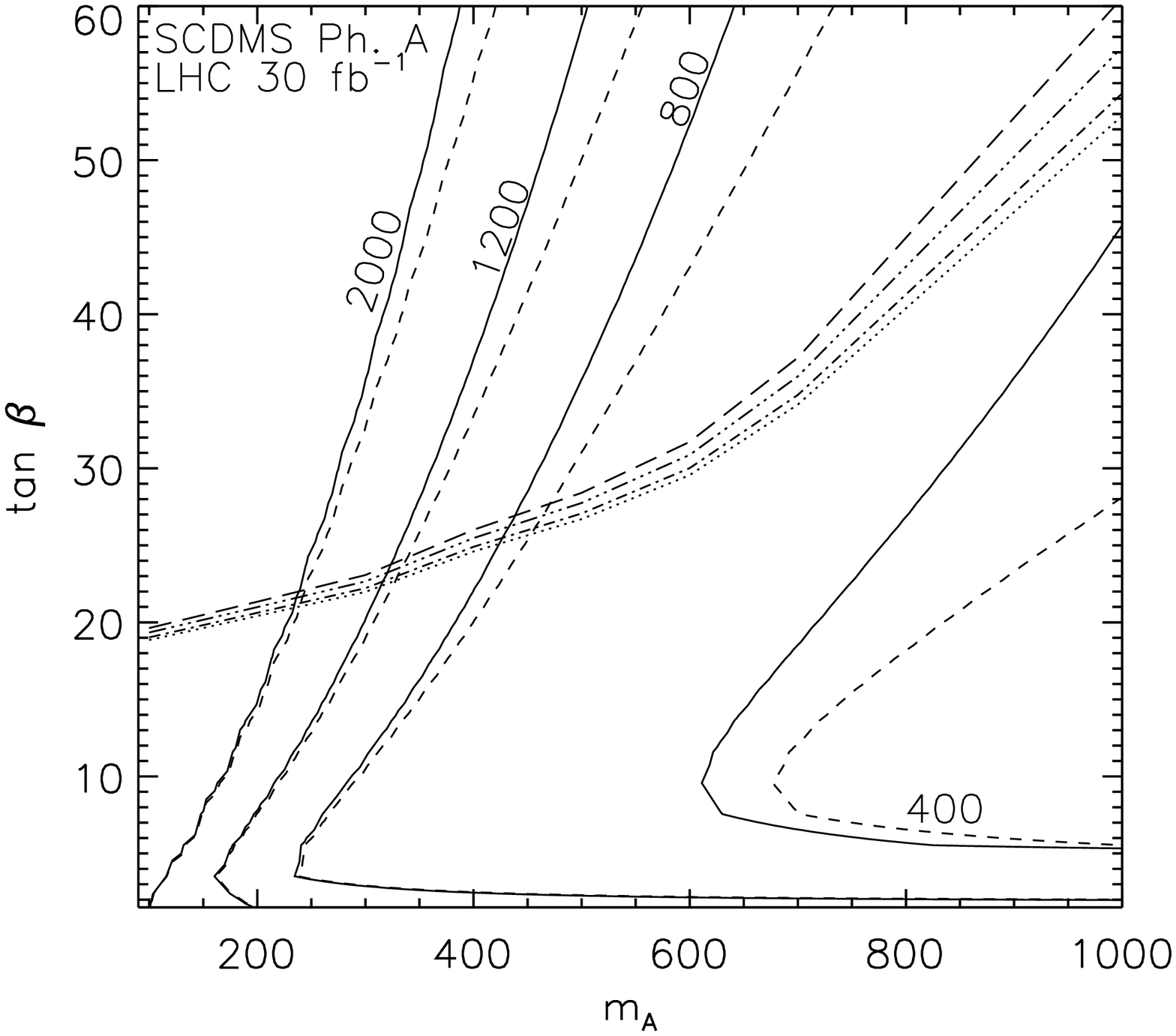}
\hspace{-1.1cm}
\includegraphics[width=3.25in,angle=0]{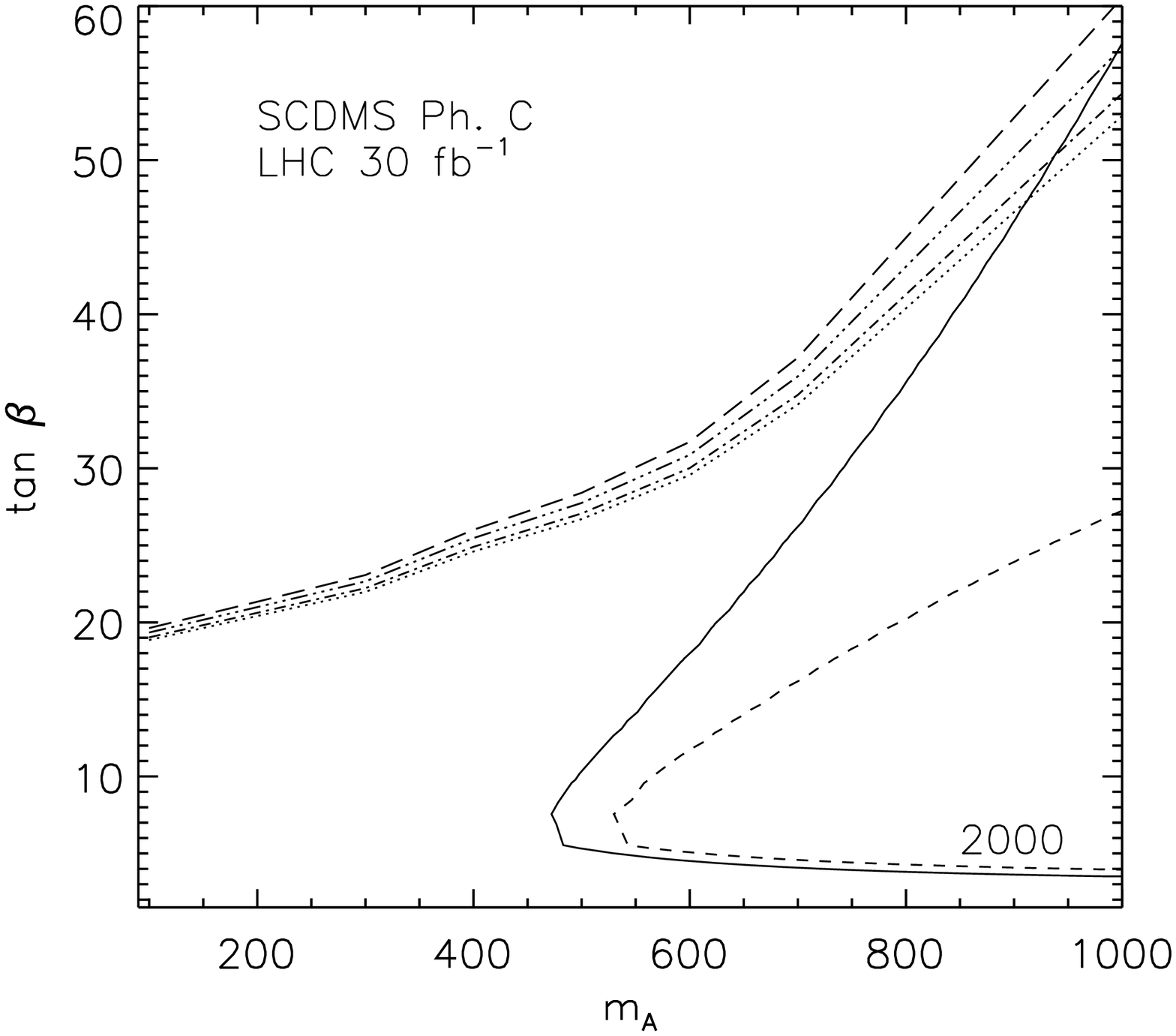}
\caption{The range of parameters in the $\tan \beta$--$m_A$ plane that can be excluded by present and future direct detection experiments (for various values of $\mu$), compared to the exclusion reach for heavy Higgs searches at the Tevatron and LHC. Regions to the left of the contours are (or will be in the future) excluded by the given experiment. The upper left frame compares the current bound from CDMS to the 1 fb$^{-1}$ exclusion region of the Tevatron. The upper right frame compares the 2007 projected limit from CDMS to the projected Tevatron limit after 4 fb$^{-1}$ integrated luminosity. The lower frames compare the projected limits of Super-CDMS (phase A and phase C in the left and right frames, respectively) to that of the LHC with 30 fb$^{-1}$. In each frame, $M_1=100$ GeV, $M_2=200$ GeV, $A_t=2 m_{\tilde{q}}+\mu \cot\beta$ (the $m^{\rm{max}}_h$ scenario) and $m_{\tilde{q}}=1$ TeV have been used. The direct detection limits are shown as solid lines which {\it do not} include contributions from squark exchange and as dotted lines which include both Higgs and squark exchange. The values of $\mu$ used for calculating the Tevatron/LHC reach are $\mu$=200 GeV (dotted), $\mu$=400 GeV (dot-dash), $\mu$=800 GeV (dot-dot-dash) and $\mu$=1200 GeV (long dashed).}
\label{compare1}
\end{figure}

\begin{figure}[!t]
\hspace{-1.0cm}
\includegraphics[width=3.25in,angle=0]{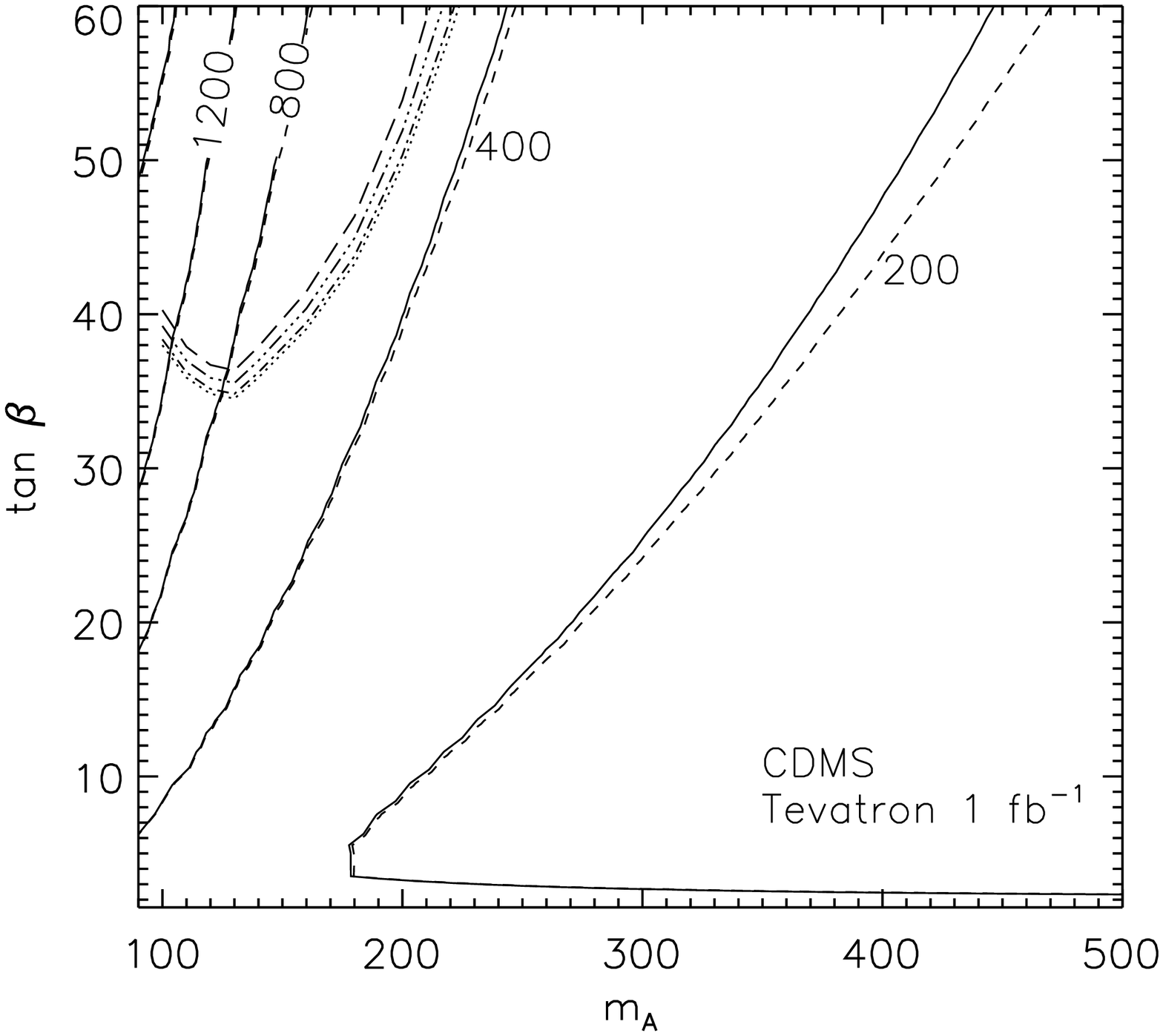}
\hspace{-1.1cm}
\includegraphics[width=3.25in,angle=0]{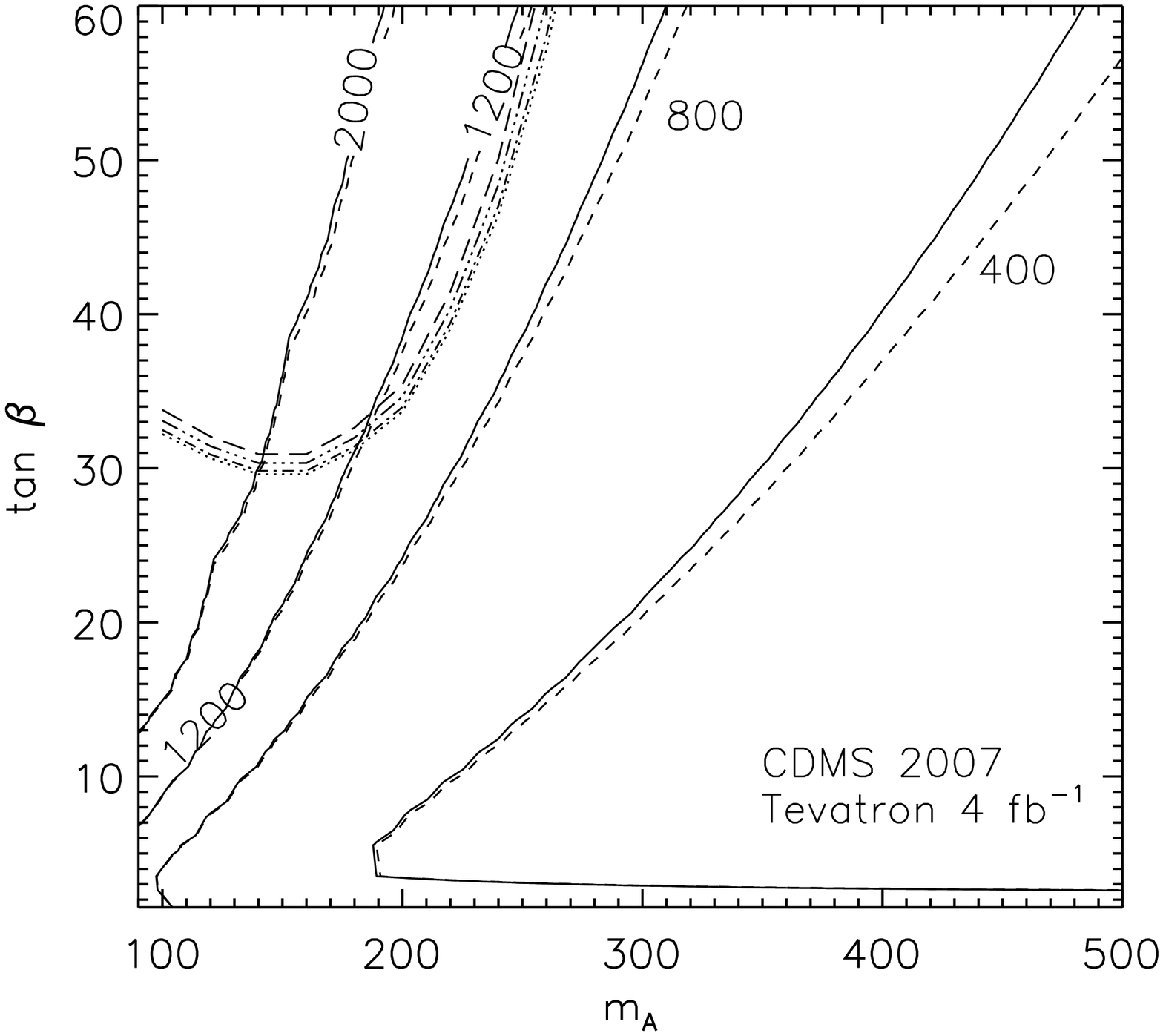}\\
\hspace{-1.0cm}
\includegraphics[width=3.25in,angle=0]{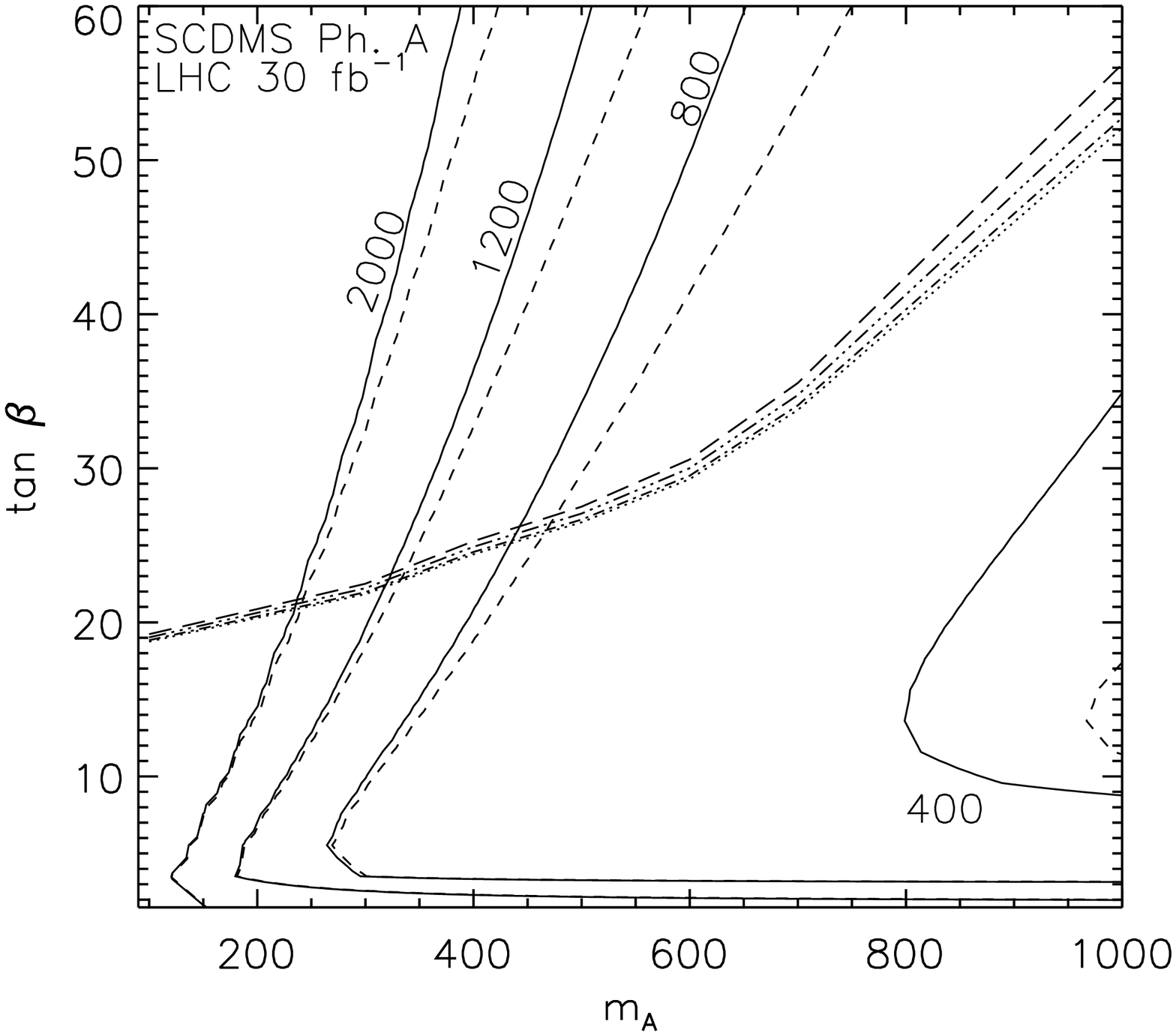}
\hspace{-1.1cm}
\includegraphics[width=3.25in,angle=0]{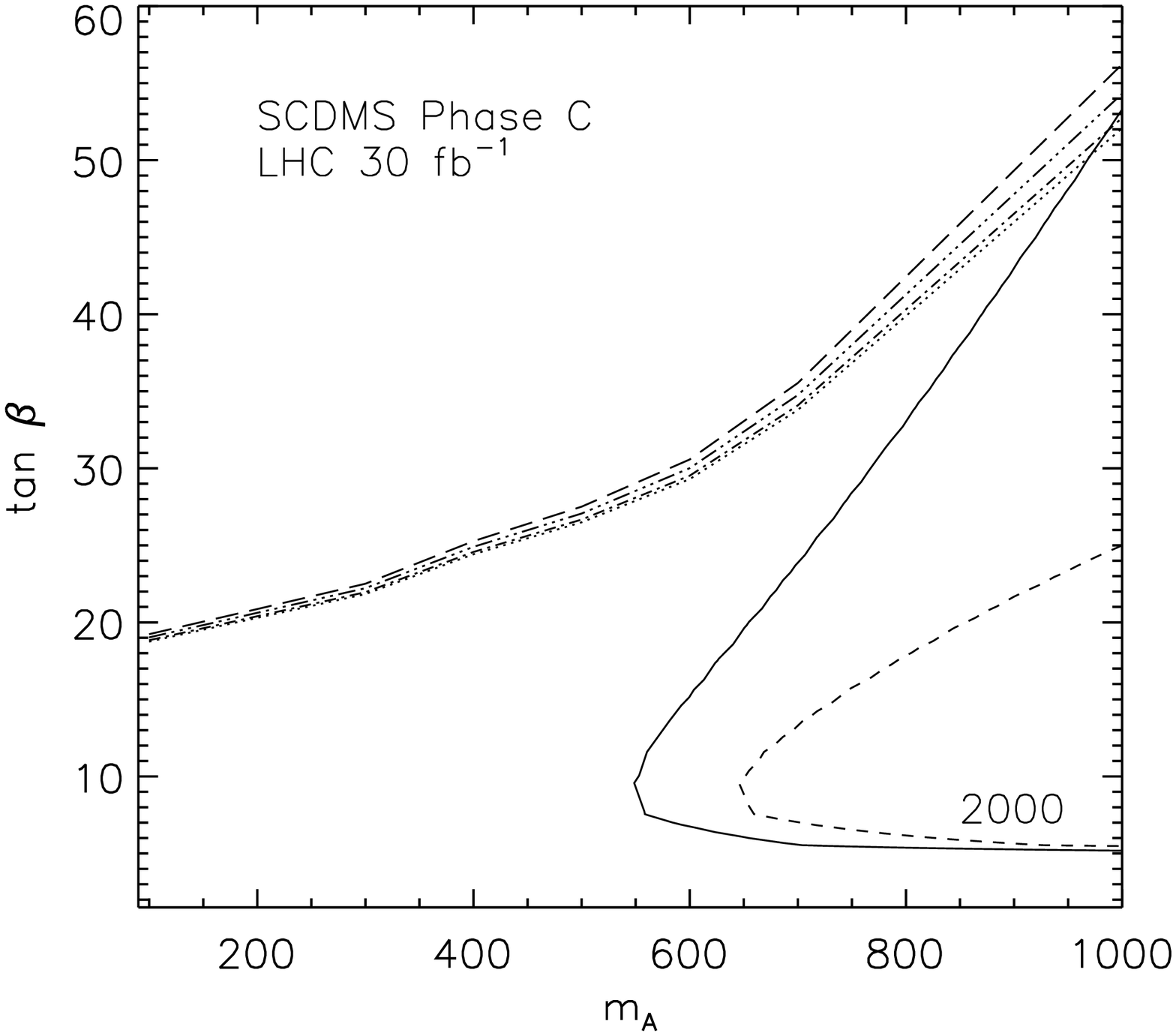}
\caption{
The same as in Fig.~\ref{compare1}, but with $A_t=\mu \cot\beta$ (the no-mixing scenario).}
\label{compare2}
\end{figure}

Since both heavy MSSM Higgs searches and neutralino direct detection depend strongly on $m_A$ and $\tan \beta$, we can compare the limits and projected reach of these experiments to each other in this plane~\cite{directsusypar}. In Figs.~\ref{compare1} and~\ref{compare2}, we show the current and projected limits from direct dark matter experiments (CDMS and Super-CDMS) along side the current and projected limits for heavy Higgs searches at the Tevatron and LHC in the $A/H +X \rightarrow \tau^+ \tau^- +X$ channel, for the $m_h^{\rm{max}}$ and no-mixing benchmark scenarios (see Sec.~\ref{susy}). From these plots, it is clear that those regions of supersymmetric parameter space most accessible to heavy Higgs searches at the Tevatron and LHC are also most likely to be probed by direct detection experiments, and vice versa. The prospects for heavy MSSM Higgs boson detection at colliders are, therefore, quite correlated to the prospects of direct detection experiments.

As can be seen in Figs.~\ref{compare1} and \ref{compare2}, the main difference in the prospects for these two types of searches is that direct detection rates depend critically on the values of $\mu$, $M_1$ and $M_2$ through their impact on the composition and mass of the lightest neutralino, while the reach of the inclusive $\tau^+ \tau^-$ channel at the Tevatron/LHC is largely independent of the precise values of these parameters. Smaller values of $\mu$ relative to $M_1$ yield larger elastic scattering cross sections and, therefore, lead to much greater sensitivity in the $m_A$-$\tan \beta$ plane. The dependance on these results of the other supersymmetric parameters are generally mild. For example, the results shown in Fig.~\ref{compare1} (the $m_h^{\rm{max}}$ scenario) are very similar to those found in Fig.~\ref{compare2} (the no-mixing scenario).

\clearpage

\section{The Implications Of Direct Detection For Colliders and Vice Versa}
\label{sec5}

As collider and direct detection experiments continue to operate, their results will have considerable implications for each other. Consider, for example, the case in which a positive detection of neutralino dark matter is made in the near future by CDMS. For models with a sufficiently large cross section to be detected by CDMS, the neutralino-nucleon scattering is typically dominanted by heavy Higgs exchange, and can be reasonably well approximated by Eq.~\ref{case1}. A large cross section therefore would be expected to favor a small value of $m_A$ and, to a lesser extent, a large value of $\tan \beta$.

Turning this scenario around, we can imagine a case in which a positive detection of the process $p\bar{p} \rightarrow A/H +X \rightarrow \tau^+ \tau^- +X$ is made in the future at the Tevatron. Such a detection would imply both a large value of $\tan \beta$ and a relatively small value of $m_A$. From Figs.~\ref{compare1} and \ref{compare2}, it is clear that if heavy Higgses are observed at the Tevatron, then CDMS is likely to observe neutralinos in the near future unless the higgsino fraction of the LSP is very small ($|\mu|$ is large). 

\begin{figure}[!t]
\hspace{-1.0cm}
\includegraphics[width=3.25in,angle=0]{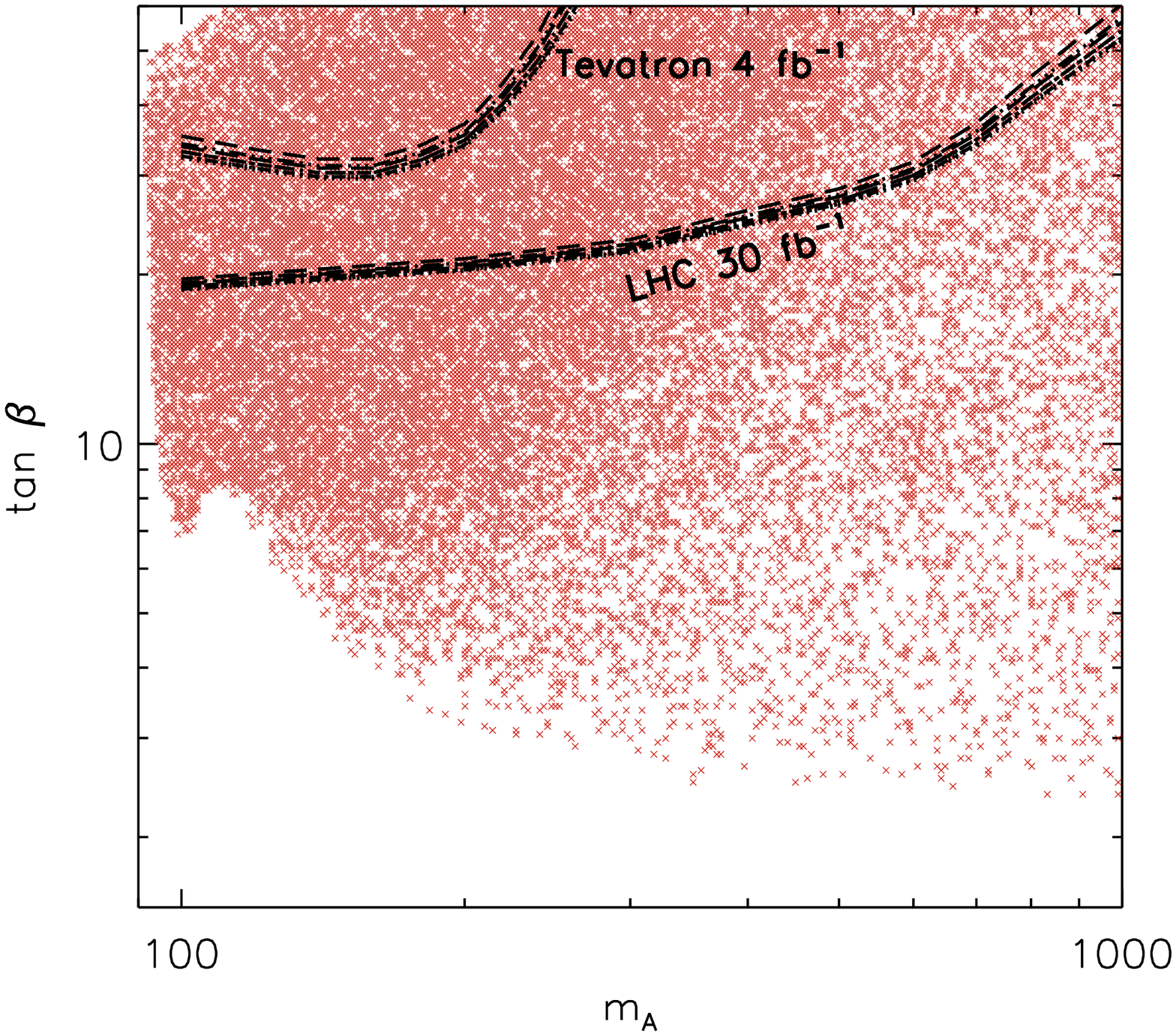}
\hspace{-1.1cm}
\includegraphics[width=3.25in,angle=0]{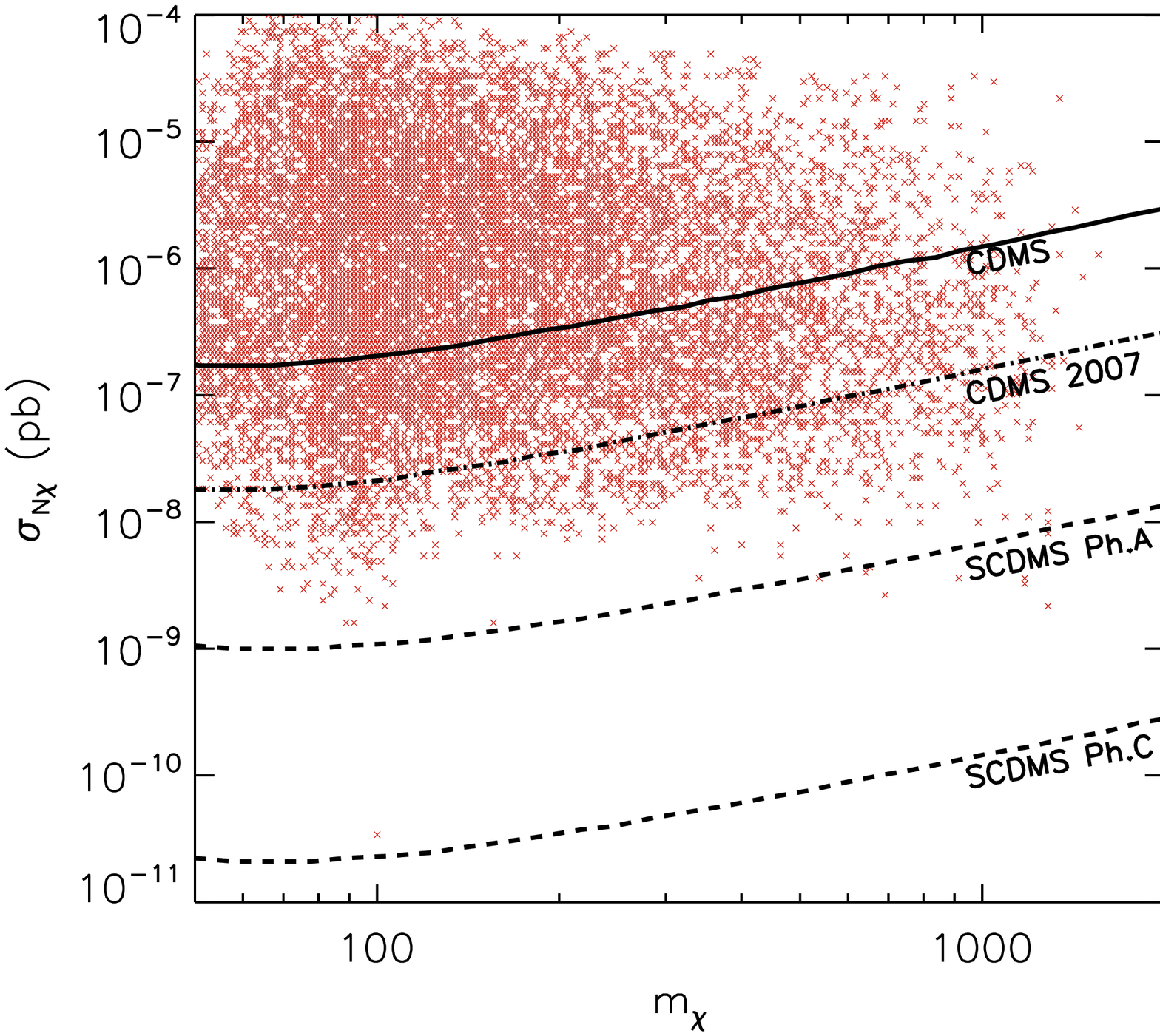}\\
\hspace{-1.0cm}
\includegraphics[width=3.25in,angle=0]{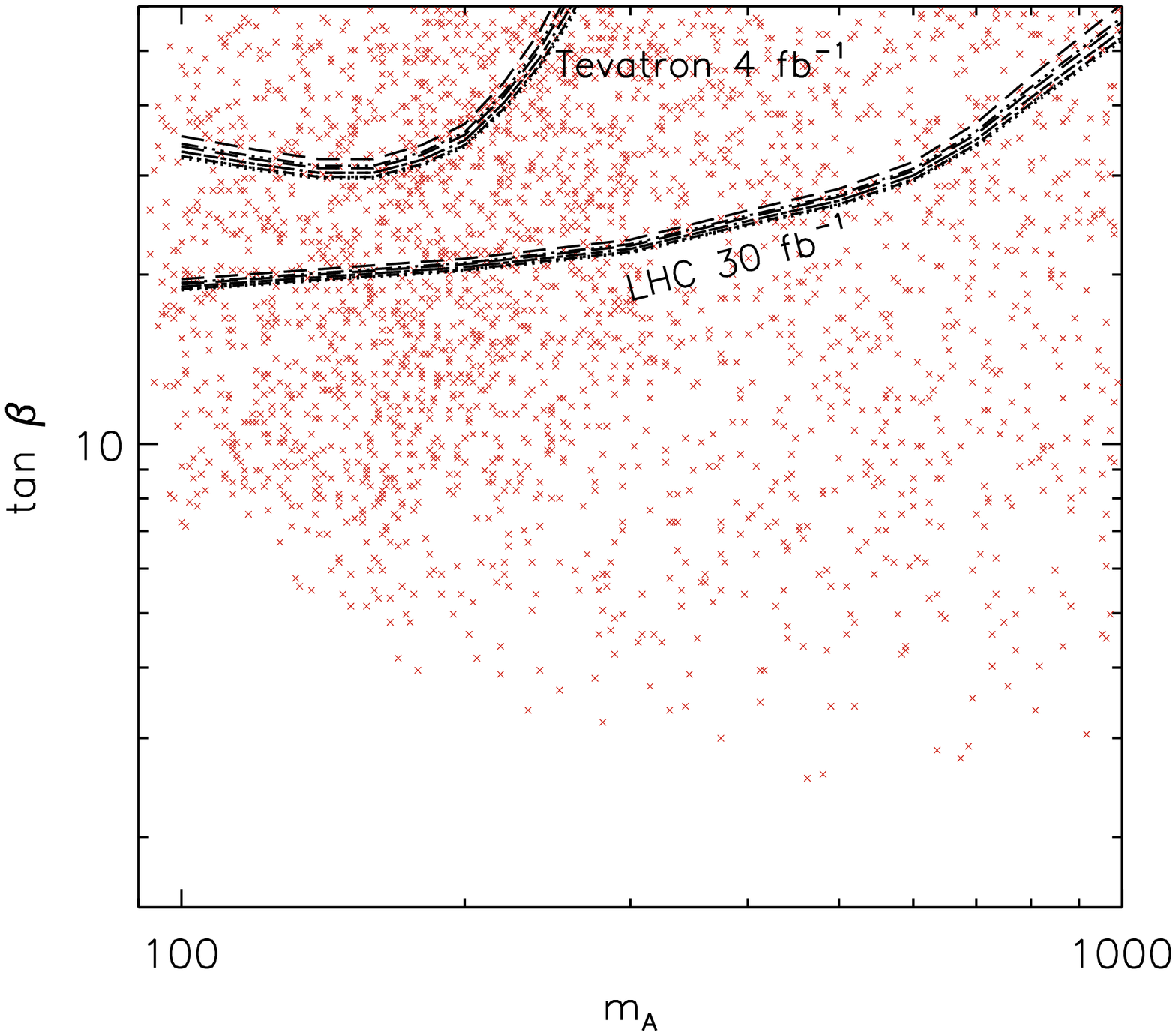}
\hspace{-1.1cm}
\includegraphics[width=3.25in,angle=0]{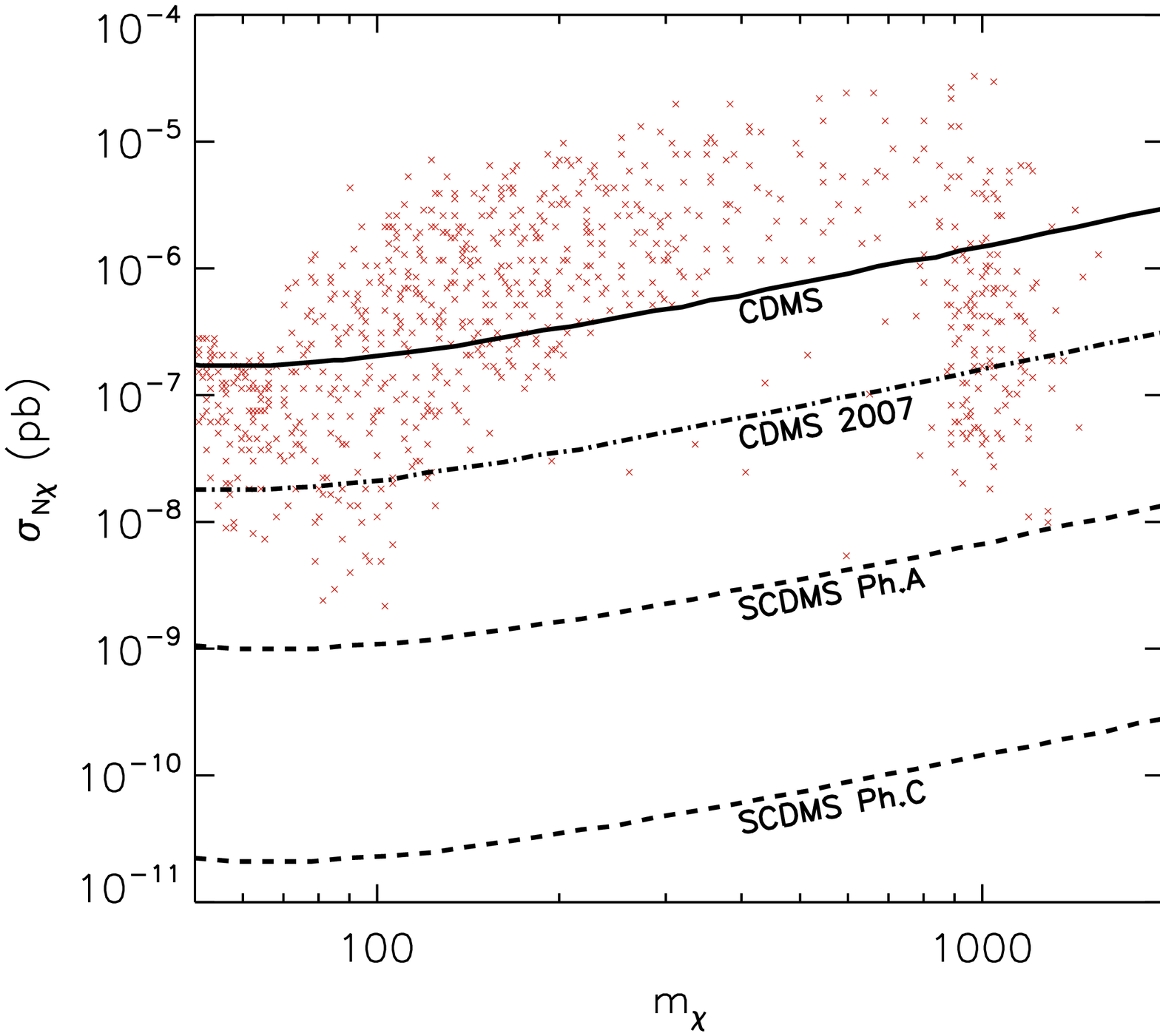}\\
\caption{Upper left: The values of $m_A$ and $\tan \beta$ for randomly selected supersymmetric models which are not currently excluded by CDMS, but are within the reach of the 2007 CDMS projection. The reach of the Tevatron and LHC heavy Higgs searches are also shown. The near future detection of neutralinos by CDMS enhances the prospects for heavy Higgs discovery at collliders, but does not guarantee any such result. Upper right:  The values of $m_{\chi}$ and $\sigma_{\chi N}$  for randomly selected supersymmetric models within the 4 fb$^{-1}$ $3\sigma$ discovery reach of the Tevatron heavy Higgs search. The 2007 projected reach of CDMS is also shown. In each frame, the points shown evade current collider constraints and provide a thermal abundance of neutralinos that does not exceed the measured dark matter density. Lower left and right frames: The same as the upper frames, but only showing those models in which the predicted thermal abundance of neutralinos matches the dark matter density as measured by WMAP ($0.131 > \Omega_{\chi} h^2 > 0.089$)~\cite{wmap}. (As opposed to only $0.131 > \Omega_{\chi} h^2$ being imposed.)}
\label{scatter1}
\end{figure}

To study the interplay between collider and direct detection searches more systematically, we have performed a scan over a range of supersymmetric parameters. In particular, we have varied $M_2$, $\mu$, $m_A$, $A_t$, $A_b$ and the sfermion masses up to 4 TeV (and with either sign of $\mu$) and $\tan \beta$ between 1 and 60. We have set the parameters $M_1$ and $M_3$ according to the conditions for gaugino mass unification. In the upper left frame of Fig.~\ref{scatter1}, we have plotted, in the $\tan$--$\beta-m_A$ plane, the models found in our scan which are not currently excluded by CDMS, but are within the reach of CDMS projected for 2007. Although models were found over most of the range of $\tan \beta$ and $m_A$ shown, the majority are concentrated at low $m_A$ and high $\tan \beta$, as expected.\footnote{Our scan was carried out logarithmically over each parameter, and therefore we have shown the results of this scan over log-scale axes in Fig.~\ref{scatter1}, in constrast to our earlier figures in the $m_A$--$\tan \beta$ plane which were presented with a linear scale.}

In the upper right frame of Fig.~\ref{scatter1} we plot, in the $m_{\chi}$-- $\sigma_{N\chi}$ plane, those models from our scan which are within the $3\sigma$ discovery reach of the Tevatron $A/H +X \rightarrow \tau^+ \tau^- +X$ search (with 4 fb$^{-1}$). As expected, we find that large elastic scattering cross sections are found for this subset of supersymmetric models. The overwhelming majority of models we find within the reach of the Tevatron are either currently excluded by CDMS or will be detected by CDMS in the near future, or by the first phase of super-CDMS.

In the upper left and upper right frames of Fig.~\ref{scatter1}, we have required that each point shown does not violate any direct collider constraints ({\it ie.} Higgs, chargino, slepton and squark mass limits) and does not predict a thermal abundance of neutralinos in excess of the dark matter density as measured by WMAP ($\Omega_{\chi} h^2 < 0.131$) \cite{wmap}. If we also require that the thermal density of neutralinos constitute the entire dark matter abundance (as opposed to non-thermal contributions being substantial), then most of the models shown in the upper frames of Fig.~\ref{scatter1} must be discarded. The lower left and lower right frames of Fig.~\ref{scatter1} are similar to the upper frames, but only include those supersymmetric models that produce a thermal relic abundance within the $3\sigma$ range measured by WMAP ($0.131 > \Omega_{\chi} h^2 > 0.089$). 

\begin{figure}[!t]
\hspace{-1.0cm}
\includegraphics[width=3.25in,angle=0]{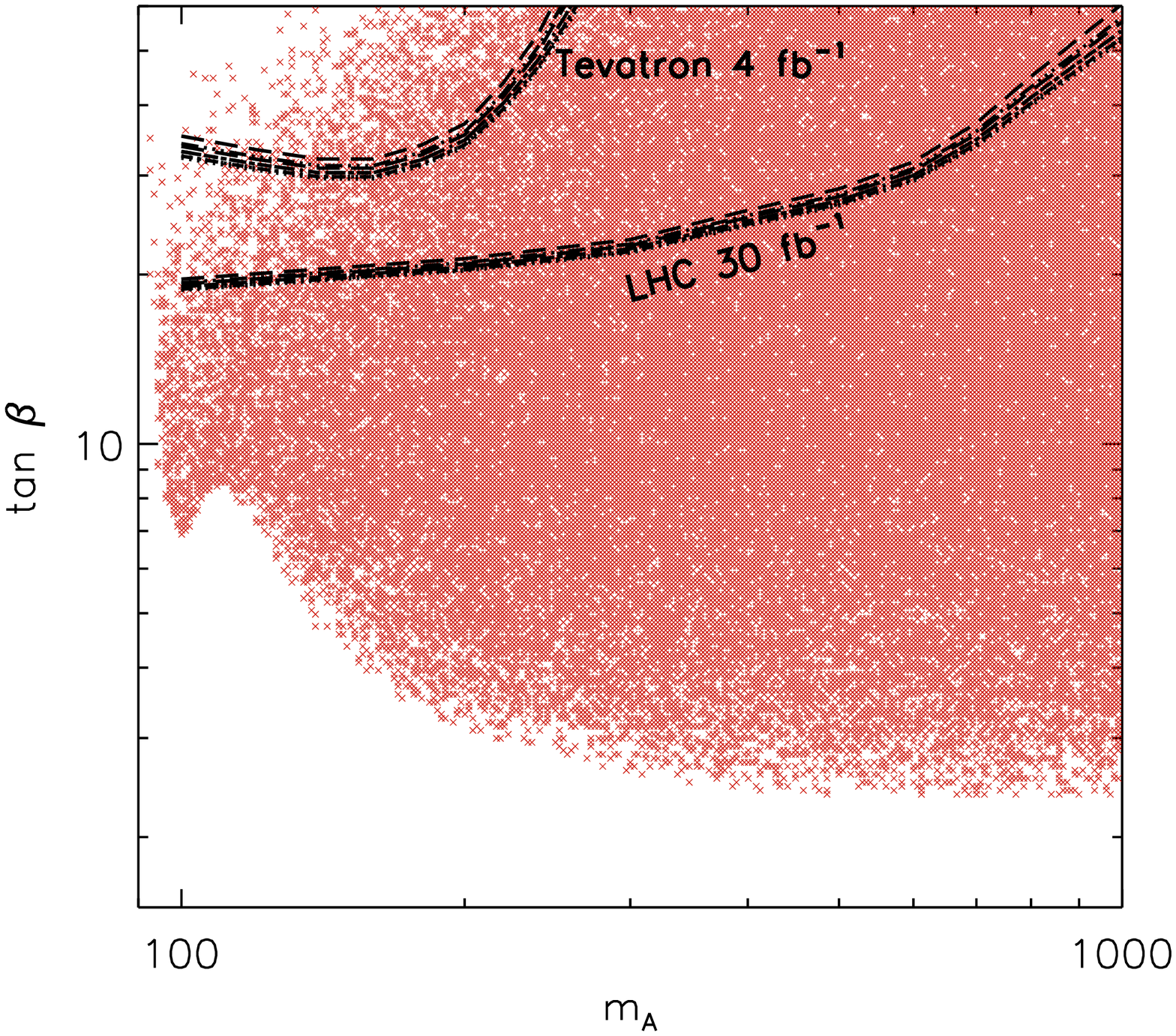}
\hspace{-1.1cm}
\includegraphics[width=3.25in,angle=0]{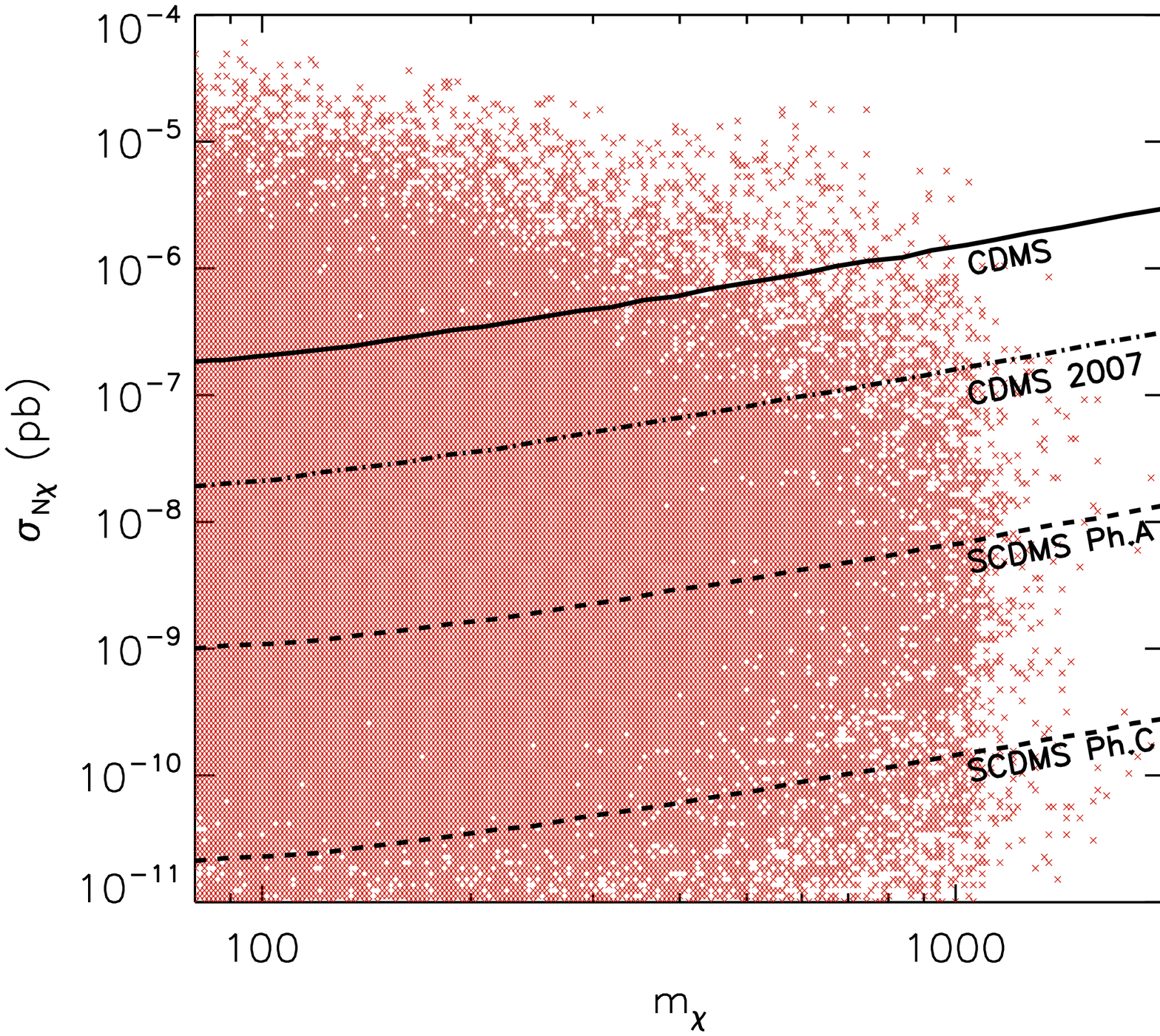}\\
\hspace{-1.0cm}
\includegraphics[width=3.25in,angle=0]{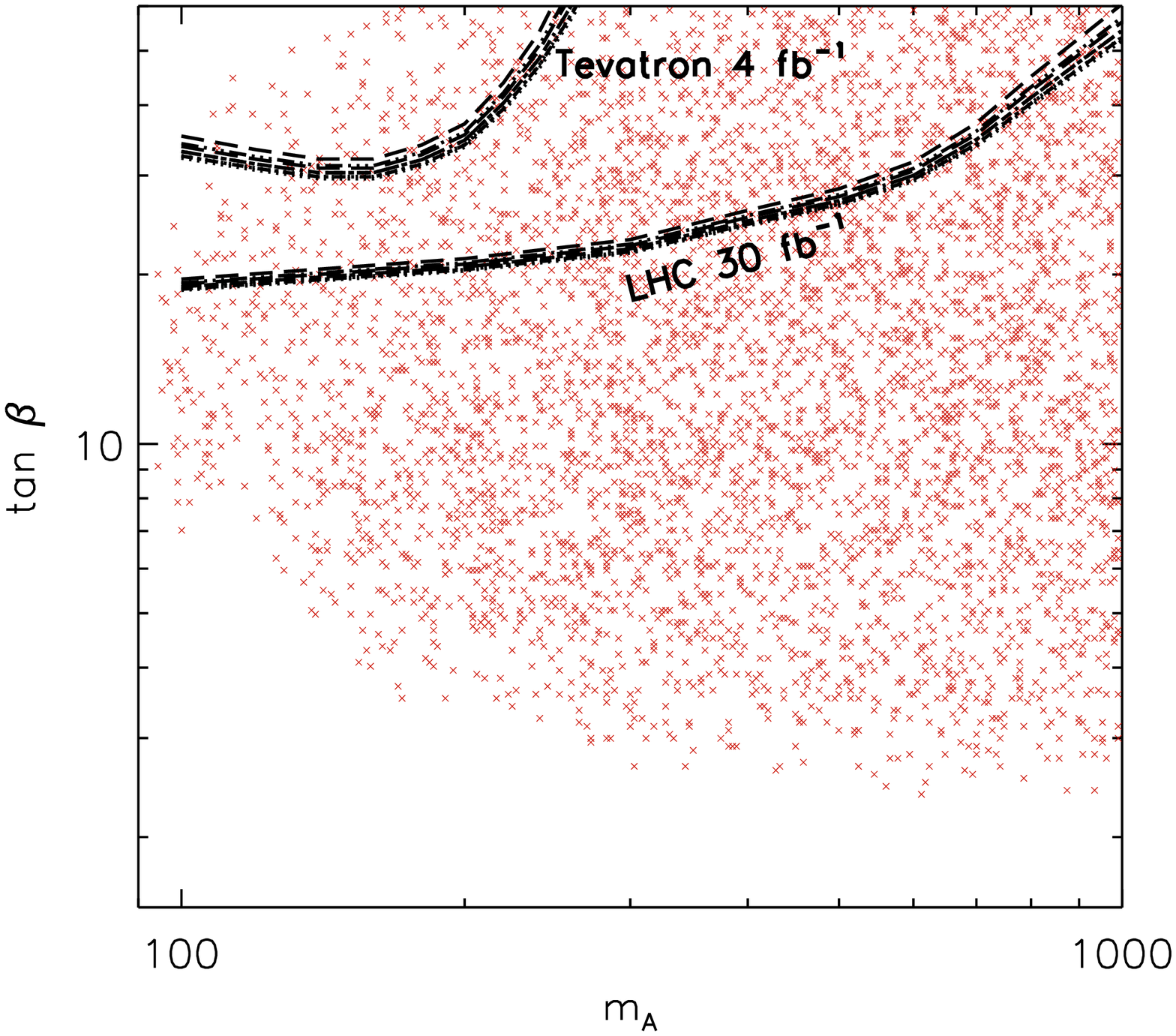}
\hspace{-1.1cm}
\includegraphics[width=3.25in,angle=0]{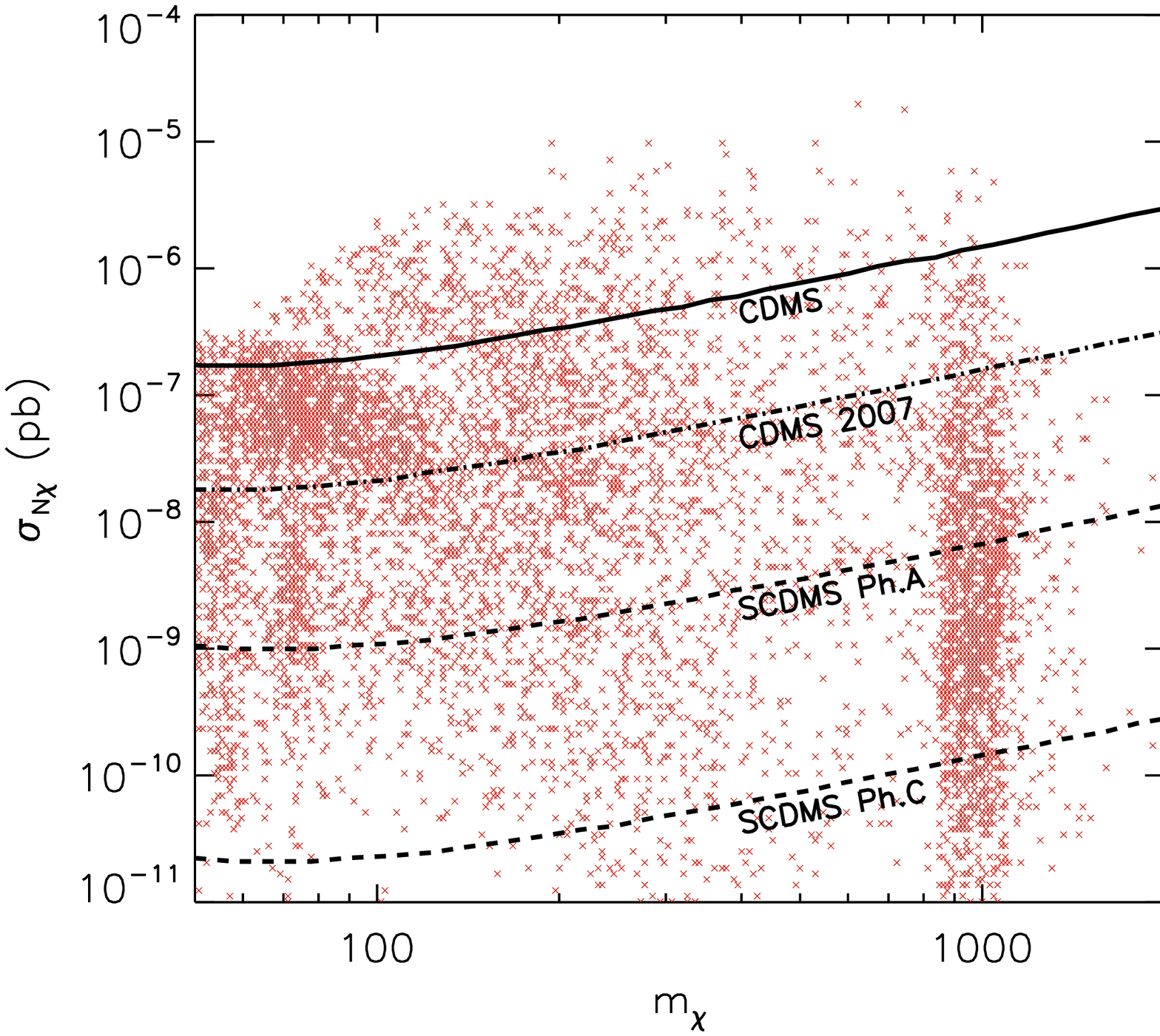}\\
\caption{The same as in Fig.~\ref{scatter1}, but showing only those supersymmetric models which are {\it not} within the 2007 projected reach of CDMS (left frames) or which are {\it not} within the projected reach of the Tevatron (after 4 fb$^{-1}$) for the process $A/H +X \rightarrow \tau^+ \tau^- +X$ (right frames). As in Fig.~\ref{scatter1}, the upper frames show those models which do not overproduce the thermal abundance of neutralino dark matter ($0.131 > \Omega_{\chi} h^2$) while the lower frames also require that the models shown do not predict a smaller thermal abundance of neutralino dark matter than measured by WMAP ($0.131 > \Omega_{\chi} h^2 > 0.089$)~\cite{wmap}.)}
\label{scatter2}
\end{figure}
\pagebreak

\begin{figure}[!t]
\hspace{-1.0cm}
\includegraphics[width=3.25in,angle=0]{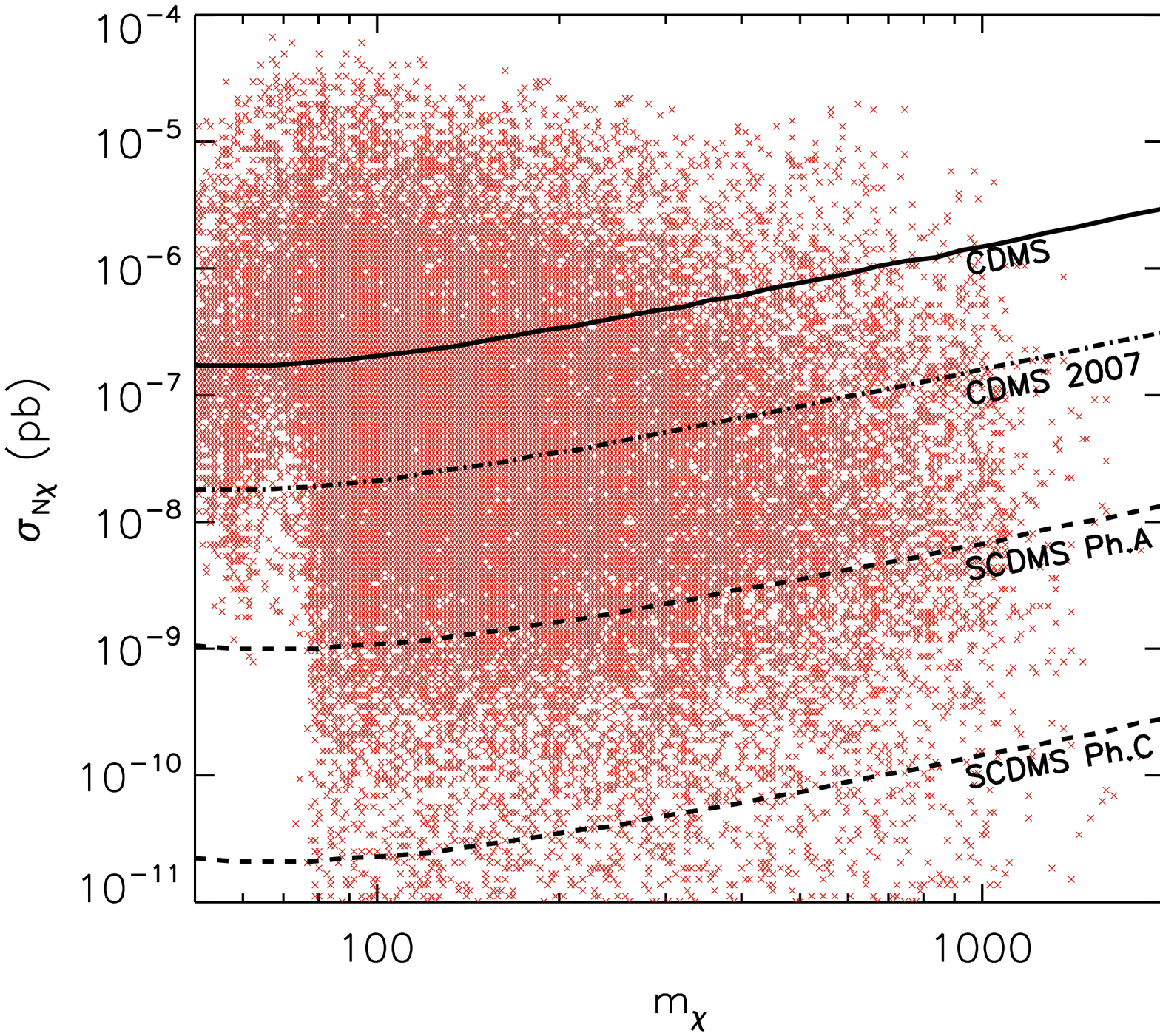}
\hspace{-1.1cm}
\includegraphics[width=3.25in,angle=0]{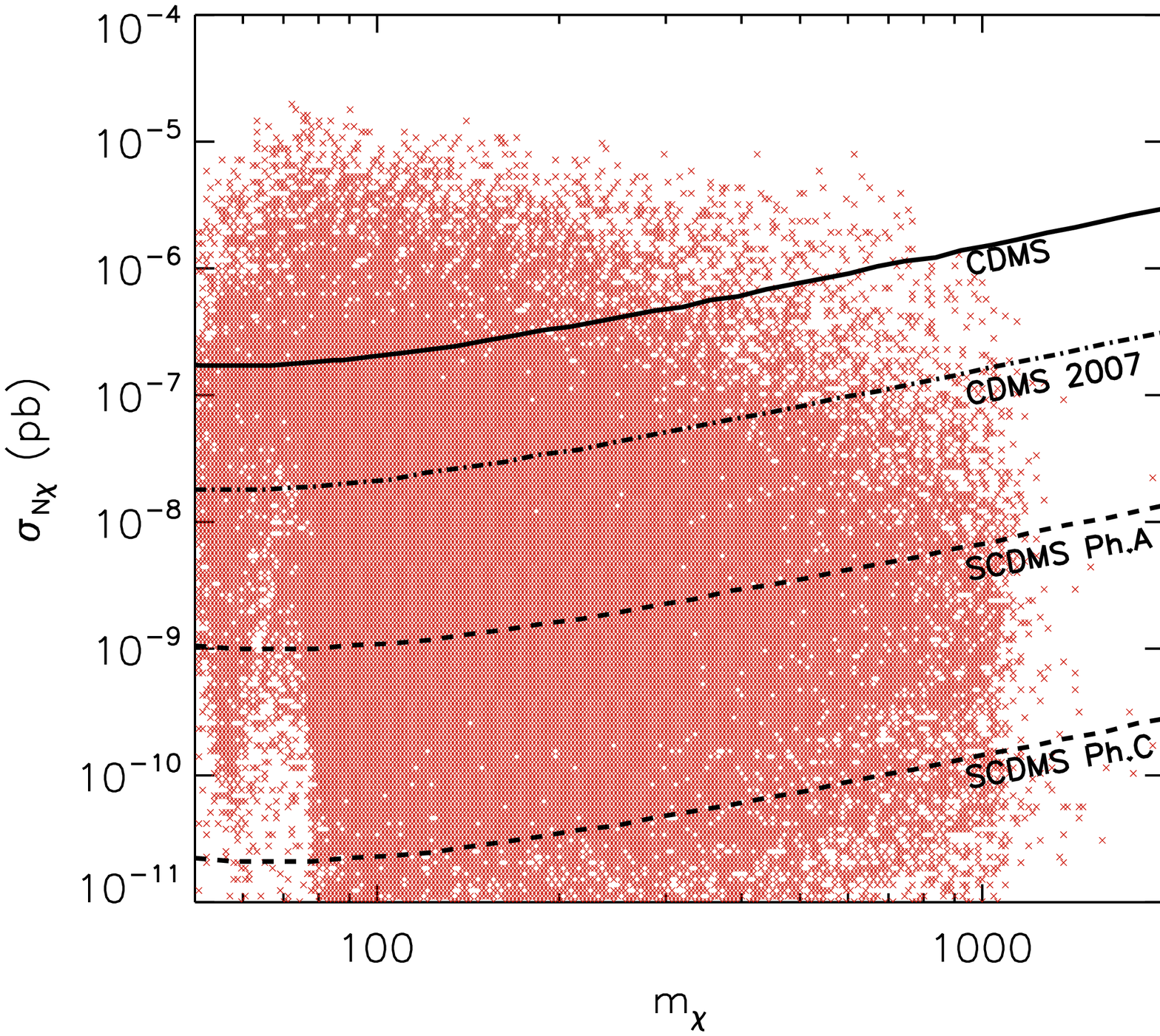}\\
\hspace{-1.0cm}
\includegraphics[width=3.25in,angle=0]{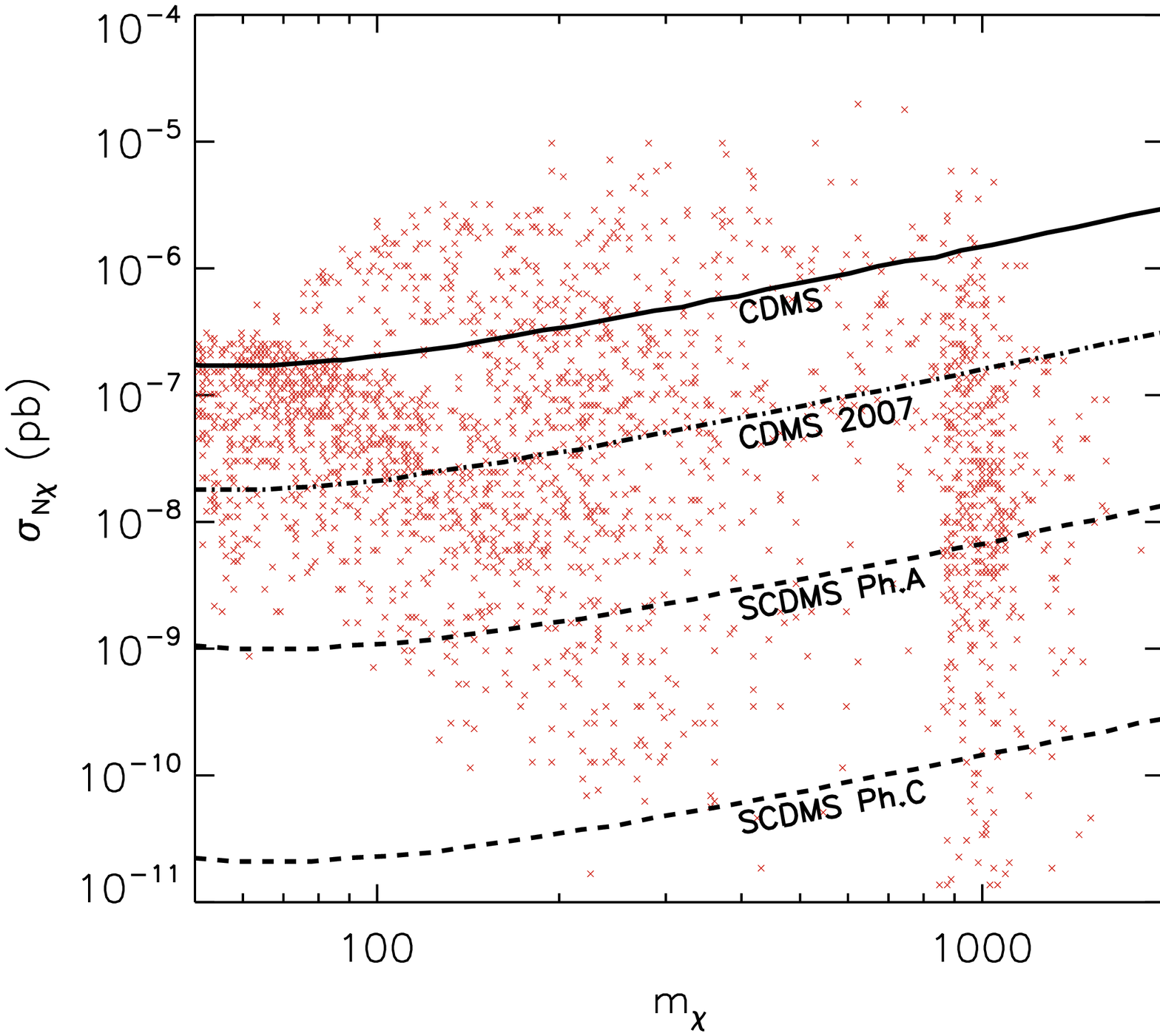}
\hspace{-1.1cm}
\includegraphics[width=3.25in,angle=0]{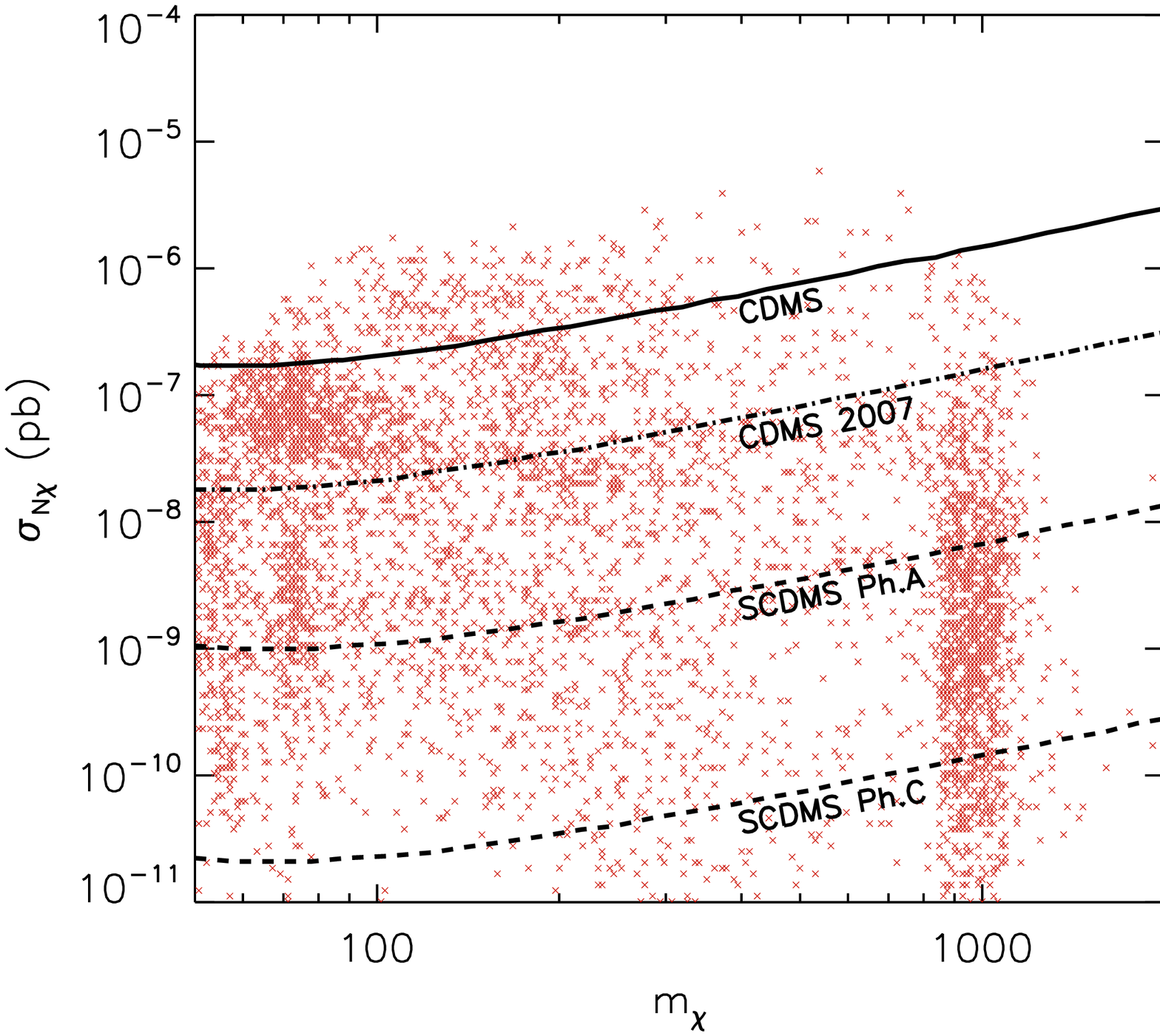}\\
\caption{The same as in the right frames of Figs.~\ref{scatter1} and \ref{scatter2}, but showing only those supersymmetric models which are (left frames) and are not (right frames) within the projected reach of the LHC (after 30 fb$^{-1}$). In each frame, those models within the reach of the Tevatron are not shown. As in Figs.~\ref{scatter1} and \ref{scatter2}, the upper frames show those models which do not overproduce the thermal abundance of neutralino dark matter ($0.131 > \Omega_{\chi} h^2$) while the lower frames also require that the models shown do not predict a smaller thermal abundance of neutralino dark matter than measured by WMAP ($0.131 > \Omega_{\chi} h^2 > 0.089$)~\cite{wmap}.)}
\label{scatterlhc}
\end{figure}
\pagebreak
If CDMS does {\it not} make a positive detection by the end of 2007, there will be implications for heavy MSSM Higgs searches at the Tevatron and LHC. Similarly, the lack of a detection of $A/H +X \rightarrow \tau^+ \tau^- +X$ at the Tevatron will impact the prospects for future direct dark matter searches. In Fig.~\ref{scatter2}, we plot (in a fashion similar to Fig.~\ref{scatter1}) the supersymmetric models found in our scan which are {\it not} within the 2007 projected reach of CDMS (left frames) or are {\it not} within the reach of the Tevatron (after 4 fb$^{-1}$) for the process $A/H +X \rightarrow \tau^+ \tau^- +X$ (right frames). If CDMS does not make a positive detection by the end of 2007, only a small fraction of the remaining models are within the reach of the Tevatron through the channel $A/H +X \rightarrow \tau^+ \tau^- +X$. These models are those in which the lightest neutralino has a very small higgsino fraction (large $|\mu|$). On the other hand, if the Tevatron does not observe the process $A/H +X \rightarrow \tau^+ \tau^- +X$, the reach of direct detection experiments can still be promising.

Whether heavy MSSM Higgs bosons are observed at the LHC will also have implications for the prospects of future direct detection experiments. In Fig.~\ref{scatterlhc} we plot, in the $\sigma_{N \chi}-m_{\chi}$ plane, those models found by our scan which are (left frames) and are not (right frames) within the reach of the LHC heavy neutral MSSM Higgs search (with 30 fb$^{-1}$). Once again, the upper figures require $0.131 > \Omega_{\chi} h^2$ while the lower frames require $0.131 > \Omega_{\chi} h^2 > 0.089$. From the left frames, we conclude that if a positive detection is made at the LHC, then the prospects for directly detecting neutralinos with CDMS or the early phases of Super-CDMS will be very promising. 

The combination of these figures demonstrates that searches at the LHC and direct dark matter searches can be highly complementary. A wide range of supersymmetric models exist in which one, both or neither of the LHC and direct detection experiments will be successful in observing heavy MSSM Higgs bosons, or neutralino dark matter, respectively.

Note that here we have only considered 30 fb$^{-1}$ of data for the LHC. Further improvements on the reach of the LHC with greater luminosity are to be expected.


\section{Uncertainties, Caveats and Limitations}
\label{sec6}

Throughout our study, we have adopted a number of assumptions. In this section, we summarize these, and discuss how our conclusions might be affected by relaxing them.

\subsection{Astrophysical Uncertainties}

Throughout our study, we have adopted a standard local dark matter density ($\rho=0.3$ GeV/cm$^3$) and velocity distribution (an isothermal Maxwell-Boltman distribution). If the actual local density or velocity distribution of dark matter is substantially different, then clearly our conclusions may be modified. 

The local density and velocity distribution of dark matter can be inferred by studying the rotation curves of our galaxy. Different groups have come to somewhat different conclusions regarding these observations: Bahcall {\it et al.} find a best-fit value of $\rho=0.34\,$GeV/cm$^3$ \cite{Bahcall:1981nv}, Caldwell and Ostriker find $\rho=0.23\,$GeV/cm$^3$ \cite{caldwellostriker} while Gates, Gyuk and Turner find $\rho=0.34-0.73\,$GeV/cm$^3$ \cite{turner}. Bergstrom, Ullio and Buckley~\cite{Bergstrom:1997fj} find that the observations are consistent with local dark matter densities in the range of about $0.2-0.8\,$GeV/cm$^3$. The uncertainties in the local dark matter velocity distribution are less important in estimating the rates in direct dark matter detection experiments \cite{kam}.

Rotation curves, however, only constrain the dark matter density as averaged over scales larger than a kiloparsec or so. In contrast, the solar system moves a distance of $\sim$$10^{-3}$ parsecs relative to the dark matter halo each year. If dark matter is distributed in an inhomogeneous way over milliparsec scales ({\it ie.} as a collection of dense clumps and voids), then the density along the path of the Earth, as seen by direct detection experiments, could be much larger or smaller than is inferred by the rotational dynamics of our galaxy. This is not anticipated to be a problem, however. The vast majority of dark matter in the inner regions of our galaxy have been in place for $\sim$$10^{10}$ years; ample time for the destruction of clumps through tidal interactions. Using high-resolution simulations, Helmi, White and Springel find that the dark matter in the solar neighborhood is likely to consist of a superposition of hundreds of thousands of dark matter streams, collectively representing a very smooth and homogeneous distribution \cite{white}.

If the uncertainties in the local halo density, the velocity distribution, and the parameters $f^{(p,n)}_T$ (from Eq.~\ref{feqn}) are taken into account, the reach of direct dark matter searches are not a single line, but instead should be thought of as a band whose width can be a factor of 3 to 5, depending on which estimates are adopted for the range of acceptable local densities.

Finally, regarding the density of dark matter, thoughout this study we have assumed that all of our universe's dark matter consists of neutralinos. If neutralinos make up only a fraction of the local dark matter density, then the rates in direct detection experiments will be reduced accordingly.

\subsection{Assumptions in the Supersymmetric Model}

As we do not know through what mechanism or mechanisms supersymmetry is broken, there is a extremely vast range of characterstics the supersymmetric spectrum might possess. To attempt to study the entire range of parameter space within the MSSM is not generally a tractable approach. If extensions of supersymmetry beyond the MSSM are considered, this becomes only more difficult. For this reason, we have adopted a number of assumptions regarding the nature of supersymmetry. In particular, we have limited ourselves to the MSSM with no CP-violating phases, and have adopted the GUT-relationship between the gaugino masses ($M_1/g_1 = M_2/g_2=M_3/g_3$). We have also adopted five common soft SUSY breaking sfermion mass parameters ($m_{Q_i}$, $m_{U_i}$, $m_{D_i}$, $m_{L_i}$ and $m_{E_i}$) at the low energy scale.

Adoping the GUT relationship between the gaugino masses guarantees that the wino-fraction of the lightest neutralino will be quite small (a few percent or less). From Eqns.~\ref{aq} and \ref{xywz}, we see that the wino fraction of the lightest neutralino, if large, can play an important role in determining its elastic scattering cross section with nucleons. In the case that $M_2 \ll M_1$, in addition to the lightest neutralino being wino-like, the lightest chargino is only slightly heavier than the LSP. In such a scenario, neutralino-neutralino annihilations and neutralino-chargino coannihilations are too efficient to produce the measured abundance of thermal neutralinos (unless $m_{\chi} \approx M_2 \gsim 3$ TeV). If all or most of the dark matter consists of neutralinos in such a model, non-thermal mechanisms must be relied upon. Non-thermally generated wino dark matter can appear naturally in models of Anomaly Mediated Supersymmetry Breaking (AMSB)~\cite{amsbdark}, for example.

We have also not considered the possibility of significant CP-violating phases in the MSSM. The presence of such phases can impact the neutralino-nucleon elastic scattering cross section, the prospects for heavy Higgs discovery at colliders, and the neutralino relic abundance through their effect on the sparticle and Higgs boson masses and couplings~\cite{cp}.

Lastly, if we consider supersymmetric models beyond the MSSM, a very broad range of possibilities become available. Neutralino dark matter has been studied in the Next-to-Minimal Supersymmetric Standard Model (NMSSM) and the near-Minimal Supersymmetric Standard Model (nMSSM) by a number of authors~\cite{nmssm}. The neutralino spectrum in these extensions contain, in addition to the four neutralinos found in the MSSM, a singlino (the superpartner of an additional Higgs singlet), leading to modified couplings and masses for the lightest neutralino. The CP-even singlet Higgs in these extensions could also mediate neutralino-nucleon elastic scattering, for example. In other extensions of the MSSM, many diverse neutralino phenomenologies can be found~\cite{barger}.

\section{Discussion and Conclusions}
\label{sec7}

In this article, we have studied the interplay between collider searches for MSSM Higgs bosons and direct dark matter experiments. In particular, the prospects for heavy MSSM Higgs searches at the Tevatron and LHC and for direct dark matter searches are each most promising in the case of large values of $\tan \beta$ and small values of $m_A$. There is, therefore, an interesting relationship between these two classes of experiments. Our findings can be summarized by the following:
\begin{itemize}

\item{If neutralinos are detected by direct dark matter experiments in the near future, then the prospects for MSSM heavy Higgs searches at the Tevatron and LHC will be significantly enhanced. In this case, however, it remains possible that the large neutralino-nucleon elastic scattering cross section could result largely from the composition of the lightest neutralino (a mixed bino-higgsino, for example) rather than from $m_A$ and $\tan \beta$ being light and large, respectively.}

\item{If the Tevatron or LHC observes heavy MSSM Higgs bosons, then the prospects for the direct detection of neutralinos will be excellent. Models which are observable at the Tevatron in these channels (with 4 fb$^{-1}$ luminosity) typically predict neutralino-nucleon cross sections in the range of $10^{-5}$ to $10^{-8}$ pb. Models which are observable at the LHC in these channels (with 30 fb$^{-1}$ luminosity) have a wide range of possible elastic scattering cross sections, roughly $10^{-5}$ to $10^{-11}$ pb, depending on the composition of the lightest neutralino.}

\item{If no WIMPs are detected by direct dark matter experiments in the near future (by CDMS by the end of 2007, for example), then MSSM heavy Higgs searches at the Tevatron are expected to be potentially successful only if the lightest neutralino is very bino-like ($|\mu| \gsim 800$ GeV).\footnote{Although Figs.~\ref{compare1} and \ref{compare2} show a value of $|\mu| \gsim 1200$ GeV rather than 800 GeV, this is the result only for the special case of $M_1=100$ GeV, as is used in those figures. The value derived from a general scan in the $M_1$-$\mu$ plane is $|\mu| \gsim 800$ GeV \cite{directsusypar}.}} The prospects for heavy MSSM Higgs discovery at the LHC are promising even if no signal is observed by direct detection experiments.

\item{If the Tevatron and/or LHC do not observe heavy MSSM Higgs bosons, then the prospects for the direct detection of neutralinos will be reduced, although many models will remain viable which are within the reach of near future and planned direct dark matter searches.}

\end{itemize}

In this work we have explored one aspect of the interplay between collider
experiments and astrophysical observations, which relies on testing
different sectors of the MSSM relevant for the understanding of dark
matter. We would like to emphasize the importance of using both collider
and direct detection experiments to study supersymmetry.
If the neutralino is discovered at the Tevatron or the LHC, it will be
through missing energy signatures in combination with jets and/or leptons.
The LSP candidate lifetime would be constrained to microsecond time scales
or longer, but the possibility that this particle is only metastable, and thus not abundant in the universe today, would
remain open.  This could occur as a result of R-parity violating couplings, for
example, or through neutralino decays to a lighter gravitino or axino \cite{superwimp}. Until the same particles are observed in terrestial direct
detection experiments, it will be impossible to know that they are stable
over cosmological time scales.

The LHC experiments will search for evidence of dark matter particles in
events with large missing energy plus multiple jets and/or leptons, as
expected in the cascades from heavy colored particles like gluinos and
squarks, for example. If the new colored particles are within the reach of
the LHC, namely if their masses are below a few TeV, then it is likely that
the LHC will find evidence for dark matter particles.  In the simplest
models analyzed, this corresponds to WIMP masses up to a few hundred GeV.  
In scenarios in which these colored particles are too heavy to be produced
at the LHC, the direct production of other new, weakly interacting
particles, which ultimately decay into the dark matter candidate, remains as a
possible search channel.  In these cases, multi-lepton signals plus
missing energy are the most robust option against the copious QCD
backgrounds.  In many case studies it appears that the properties and
couplings of the new particles can be measured with, at best, only modest
precision, and hence it will be difficult to establish the identity of the
dark matter particle from LHC data alone. Direct dark matter detection experiments would open another
window into the nature and composition of neutralinos, and hence play an
essential and complementary role in the quantitative study of supersymmetric dark matter~\cite{Hooper:2006wv}.

In the more distant future, experiments at the prospective International
Linear Collider (ILC) would measure many of the properties of
supersymmetric particles much more accurately than can be done at the LHC
\cite{ilc}. In the case that the non-strongly interacting particles associated with the dark matter are within the kinematic reach of
the ILC, these particles can be produced and their masses and couplings
measured with high precision. The availability of polarized beams and the
capability to make precise measurements of cross sections is particularly
useful. The ILC can also give direct information on particle masses from
the kinematic distributions of decay products and from the measurement of
excitation curves at center-of-mass energies near threshold.  In the case
of heavy new particles that are weakly interacting, the above measurements
can constrain the heavy masses and couplings relevant to the computation
of the dark matter relic density.  Ultimately, information from the
ILC can determine the relevant supersymmetric parameters
with sufficient accuracy such that the relic density can be computed to the percent level and compared with cosmological measurements of similar
precision. Given an understanding of the properties of the neutralino at
this level, measurements of rates at direct (and indirect)  
dark matter detection experiments will allow one to infer the local
distribution of dark matter \cite{baltz,freitas}, and even constrain
the expansion history of our universe since the time of neutralino
freeze-out.

\acknowledgements{DH and AV are supported by the Department of Energy and by NASA grant NAG5-10842. MC is supported by the US Department of Energy grant DE-AC02-76CHO3000.}


\begin{thebibliography}{10}






\bibitem{cdms}
 D.~S.~Akerib {\it et al.}  [CDMS Collaboration],
  Phys.\ Rev.\ Lett.\  {\bf 96}, 011302 (2006)
  [arXiv:astro-ph/0509259];
 D.~S.~Akerib {\it et al.}  [CDMS Collaboration],
  Phys.\ Rev.\ D {\bf 73}, 011102 (2006)
  [arXiv:astro-ph/0509269].

\bibitem{zeplin}
  G.~J.~Alner {\it et al.}  [UK Dark Matter Collaboration],
  Astropart.\ Phys.\  {\bf 23}, 444 (2005).

\bibitem{edelweiss}
  V.~Sanglard {\it et al.}  [The EDELWEISS Collaboration],
  Phys.\ Rev.\ D {\bf 71}, 122002 (2005)
  [arXiv:astro-ph/0503265].

\bibitem{cresst}
  G.~Angloher {\it et al.},
  Astropart.\ Phys.\  {\bf 23}, 325 (2005)
  [arXiv:astro-ph/0408006].



\bibitem{warp}
  R.~Brunetti {\it et al.},
  New Astron.\ Rev.\  {\bf 49}, 265 (2005)
  [arXiv:astro-ph/0405342];
Preliminary limits using liquid Argon were shown at the April 2006, P5 meeting at Fermilab by Elena Aprile (http://www.fnal.gov/directorate/program$_{-}$planning/P5/P5$_{-}$Apr2006/Talks/Aprile.pdf).


\bibitem{dmreview}
  G.~Bertone, D.~Hooper and J.~Silk,
  Phys.\ Rept.\  {\bf 405}, 279 (2005)
  [arXiv:hep-ph/0404175].

\bibitem{tevneucha}
  D.~Bortoletto  [CDF and D0 Collaborations],
  PoS {\bf HEP2005}, 347 (2006);
  A.~Canepa  [CDF Collaboration],
  arXiv:hep-ex/0603032.

\bibitem{tevhiggs}
  A.~Anastassov  [CDF and D0 Collaborations],
  PoS {\bf HEP2005}, 326 (2006);

\bibitem{tevsquarksgluinos}
  V.~Abazov  [D0 Collaboration],
  arXiv:hep-ex/0604029.



\bibitem{directsusypar}
  M.~Carena, D.~Hooper and P.~Skands,
Phys.\ Rev.\ Lett.\  {\bf 97}, 051801 (2006)
  [arXiv:hep-ph/0603180].



\bibitem{maxnomixing}
  M.~Carena, S.~Heinemeyer, C.~E.~M.~Wagner and G.~Weiglein,
  arXiv:hep-ph/9912223.

\bibitem{feynhiggs}
 S.~Heinemeyer, W.~Hollik and G.~Weiglein,
  Comput.\ Phys.\ Commun.\  {\bf 124}, 76 (2000)
  [arXiv:hep-ph/9812320].

\bibitem{cpsuperh}
  J.~S.~Lee, A.~Pilaftsis, M.~Carena, S.~Y.~Choi, M.~Drees, J.~R.~Ellis and C.~E.~M.~Wagner,
  Comput.\ Phys.\ Commun.\  {\bf 156}, 283 (2004)
  [arXiv:hep-ph/0307377].


\bibitem{primer}
  S.~P.~Martin,
  arXiv:hep-ph/9709356.

\bibitem{susyreview2}
  M.~Carena, R.~L.~Culbertson, S.~Eno, H.~J.~Frisch and S.~Mrenna,
  Rev.\ Mod.\ Phys.\  {\bf 71}, 937 (1999)
  [arXiv:hep-ex/9712022].

\bibitem{higgsreview}
  M.~Carena and H.~E.~Haber,
  Prog.\ Part.\ Nucl.\ Phys.\  {\bf 50}, 63 (2003)
  [arXiv:hep-ph/0208209].

\bibitem{nuc}
  A.~Bottino, F.~Donato, N.~Fornengo and S.~Scopel,
  Astropart.\ Phys.\  {\bf 18}, 205 (2002)
  [arXiv:hep-ph/0111229];
 Astropart.\ Phys.\  {\bf 13}, 215 (2000)
  [arXiv:hep-ph/9909228];
  J.~R.~Ellis, K.~A.~Olive, Y.~Santoso and V.~C.~Spanos,
  Phys.\ Rev.\ D {\bf 71}, 095007 (2005)
  [arXiv:hep-ph/0502001].

\bibitem{scatteraq}
G.~B.~Gelmini, P.~Gondolo and E.~Roulet,
Nucl.\ Phys.\ B {\bf 351}, 623 (1991);
M.~Srednicki and R.~Watkins,
Phys.\ Lett.\ B {\bf 225}, 140 (1989);
M.~Drees and M.~Nojiri,
Phys.\ Rev.\ D {\bf 48}, 3483 (1993)
[arXiv:hep-ph/9307208];
M.~Drees and M.~M.~Nojiri,
Phys.\ Rev.\ D {\bf 47}, 4226 (1993)
[arXiv:hep-ph/9210272];
J.~R.~Ellis, A.~Ferstl and K.~A.~Olive, 
Phys.~Lett.~B  481, (2000) 304,
[arXiv:hep-ph/0001005].



\bibitem{higgscdfcurrent}
  A.~Abulencia {\it et al.}  [CDF Collaboration],
  Phys.\ Rev.\ Lett.\  {\bf 96}, 011802 (2006)
  [arXiv:hep-ex/0508051];
 A.~Abulencia {\it et al.}  [CDF Collaboration],
  Phys.\ Rev.\ Lett.\  {\bf 96}, 042003 (2006)
  [arXiv:hep-ex/0510065].

\bibitem{higgsdzerocurrent}
  V.~M.~Abazov {\it et al.}  [D0 Collaboration],
  arXiv:hep-ex/0605009;
  V.~M.~Abazov {\it et al.}  [D0 Collaboration],
  Phys.\ Rev.\ Lett.\  {\bf 95}, 151801 (2005)
  [arXiv:hep-ex/0504018];
www-d0.fnal.gov/Run2Physics/WWW/results/prelim/HIGGS/H24/H24.pdf.

\bibitem{carena}
  M.~Carena, S.~Heinemeyer, C.~E.~M.~Wagner and G.~Weiglein,
  Eur.\ Phys.\ J.\ C {\bf 45}, 797 (2006)
  [arXiv:hep-ph/0511023].



\bibitem{gminus2}
 G.~W.~Bennett {\it et al.}  [Muon g-2 Collaboration],
  Phys.\ Rev.\ Lett.\  {\bf 92}, 161802 (2004)
  [arXiv:hep-ex/0401008];
  R.~R.~Akhmetshin {\it et al.}  [CMD-2 Collaboration],
  Phys.\ Lett.\ B {\bf 578}, 285 (2004)
  [arXiv:hep-ex/0308008];
  M.~Davier, S.~Eidelman, A.~Hocker and Z.~Zhang,
  Eur.\ Phys.\ J.\ C {\bf 31}, 503 (2003)
  [arXiv:hep-ph/0308213];
  F.~Jegerlehner,
  Nucl.\ Phys.\ Proc.\ Suppl.\  {\bf 126}, 325 (2004)
  [arXiv:hep-ph/0310234];
 K.~Hagiwara, A.~D.~Martin, D.~Nomura and T.~Teubner,
  Phys.\ Rev.\ D {\bf 69}, 093003 (2004)
  [arXiv:hep-ph/0312250].

\bibitem{gminus2theory}
  T.~Moroi,
  Phys.\ Rev.\ D {\bf 53}, 6565 (1996)
  [Erratum-ibid.\ D {\bf 56}, 4424 (1997)]
  [arXiv:hep-ph/9512396];
  M.~Carena, G.~F.~Giudice and C.~E.~M.~Wagner,
  Phys.\ Lett.\ B {\bf 390}, 234 (1997)
  [arXiv:hep-ph/9610233];
  L.~L.~Everett, G.~L.~Kane, S.~Rigolin and L.~T.~Wang,
  Phys.\ Rev.\ Lett.\  {\bf 86}, 3484 (2001)
  [arXiv:hep-ph/0102145];
  E.~A.~Baltz and P.~Gondolo,
  Phys.\ Rev.\ D {\bf 67}, 063503 (2003)
  [arXiv:astro-ph/0207673].

\bibitem{bsg}
 M.~S.~Alam {\it et al.}  [CLEO Collaboration],
  Phys.\ Rev.\ Lett.\  {\bf 74}, 2885 (1995);
 K.~Abe {\it et al.}  [Belle Collaboration], 
  arXiv:hep-ex/0107065;
 L.~Lista  [BABAR Collaboration],
  arXiv:hep-ex/0110010;
 B.~Aubert {\it et al.}  [BABAR Collaboration],
  arXiv:hep-ex/0207074;

\bibitem{bsgtheory}
  R.~Barbieri and G.~F.~Giudice,
  Phys.\ Lett.\ B {\bf 309}, 86 (1993)
  [arXiv:hep-ph/9303270];
  S.~Bertolini, F.~Borzumati, A.~Masiero and G.~Ridolfi,
  Nucl.\ Phys.\ B {\bf 353}, 591 (1991);
  M.~Carena, D.~Garcia, U.~Nierste and C.~E.~M.~Wagner,
  Phys.\ Lett.\ B {\bf 499}, 141 (2001)
  [arXiv:hep-ph/0010003];
  G.~Degrassi, P.~Gambino and G.~F.~Giudice,
  JHEP {\bf 0012}, 009 (2000)
  [arXiv:hep-ph/0009337];
  H.~Baer, M.~Brhlik, D.~Castano and X.~Tata,
  Phys.\ Rev.\ D {\bf 58}, 015007 (1998)
  [arXiv:hep-ph/9712305];
  M.~Carena, A.~Menon, R.~Noriega-Papaqui, A.~Szynkman and C.~E.~M.~Wagner,
  Phys.\ Rev.\ D {\bf 74}, 015009 (2006)
  [arXiv:hep-ph/0603106].



\bibitem{cdfexotics}
CDF exotics public webpage:  http://www-cdf.fnal.gov/physics/exotic/r2a/20050519.mssm$_{-}$htt/\\projections.htm


\bibitem{darksusy}
 P.~Gondolo, J.~Edsjo, P.~Ullio, L.~Bergstrom, M.~Schelke and E.~A.~Baltz,
  New Astron.\ Rev.\  {\bf 49}, 149 (2005);
  P.~Gondolo, J.~Edsjo, P.~Ullio, L.~Bergstrom, M.~Schelke and E.~A.~Baltz,
  JCAP {\bf 0407}, 008 (2004)
  [arXiv:astro-ph/0406204].


\bibitem{wmap}
D.~N.~Spergel {\it et al}. [WMAP Collaboration],
arXiv:astro-ph/0603449.



\bibitem{Bahcall:1981nv}
J.~N.~Bahcall, M.~Schmidt and R.~M.~Soneira,
Astrophys.~J.~265 (1983) 730.

\bibitem{caldwellostriker}
R.~R.~Caldwell and J.~P.~Ostriker, 
Astrophys.~J.~251 (1981) 61. 

\bibitem{turner}
E.~I.~Gates, G.~Gyuk and M.~S.~Turner,
Astrophys.\ J.\ {\bf 449}, L123 (1995)
[astro-ph/9505039].


\bibitem{Bergstrom:1997fj}
L.~Bergstrom, P.~Ullio and J.~H.~Buckley,
Astropart.\ Phys.\  {\bf 9} (1998) 137
[arXiv:astro-ph/9712318].

\bibitem{kam}
M.~Kamionkowski and A.~Kinkhabwala,
Phys.\ Rev.\ D {\bf 57}, 3256 (1998)
[hep-ph/9710337];




\bibitem{white}
  A.~Helmi, S.~D.~M.~White and V.~Springel,
  Phys.\ Rev.\ D {\bf 66}, 063502 (2002)
  [arXiv:astro-ph/0201289].



\bibitem{amsbdark}
T.~Moroi and L.~Randall,
Nucl.\ Phys.\ B {\bf 570}, 455 (2000)
[arXiv:hep-ph/9906527];
 P.~Ullio,
  JHEP {\bf 0106}, 053 (2001)
  [arXiv:hep-ph/0105052];
  D.~Hooper and L.~T.~Wang,
  Phys.\ Rev.\ D {\bf 69}, 035001 (2004)
  [arXiv:hep-ph/0309036].






\bibitem{cp}
  C.~Balazs, M.~Carena and C.~E.~M.~Wagner,
  Phys.\ Rev.\ D {\bf 70}, 015007 (2004)
  [arXiv:hep-ph/0403224];
 C.~Balazs, M.~Carena, A.~Menon, D.~E.~Morrissey and C.~E.~M.~Wagner,
  Phys.\ Rev.\ D {\bf 71}, 075002 (2005)
  [arXiv:hep-ph/0412264];
  G.~Belanger, F.~Boudjema, S.~Kraml, A.~Pukhov and A.~Semenov,
  Phys.\ Rev.\ D {\bf 73}, 115007 (2006)
  [arXiv:hep-ph/0604150];
  T.~Nihei and M.~Sasagawa,
  Phys.\ Rev.\ D {\bf 70}, 055011 (2004)
  [Erratum-ibid.\ D {\bf 70}, 079901 (2004)]
  [arXiv:hep-ph/0404100].

\bibitem{nmssm}
  D.~G.~Cerdeno, C.~Hugonie, D.~E.~Lopez-Fogliani, C.~Munoz and A.~M.~Teixeira,
  JHEP {\bf 0412}, 048 (2004)
  [arXiv:hep-ph/0408102];
  G.~Belanger, F.~Boudjema, C.~Hugonie, A.~Pukhov and A.~Semenov,
  JCAP {\bf 0509}, 001 (2005)
  [arXiv:hep-ph/0505142];
  J.~F.~Gunion, D.~Hooper and B.~McElrath,
  Phys.\ Rev.\ D {\bf 73}, 015011 (2006)
  [arXiv:hep-ph/0509024];
  V.~Barger, P.~Langacker and G.~Shaughnessy,
  arXiv:hep-ph/0609068;
  A.~Menon, D.~E.~Morrissey and C.~E.~M.~Wagner,
  Phys.\ Rev.\ D {\bf 70}, 035005 (2004)
  [arXiv:hep-ph/0404184].



\bibitem{barger}
  V.~Barger, P.~Langacker and H.~S.~Lee,
  Phys.\ Lett.\ B {\bf 630}, 85 (2005)
  [arXiv:hep-ph/0508027];
  V.~Barger, C.~Kao, P.~Langacker and H.~S.~Lee,
  Phys.\ Lett.\ B {\bf 600}, 104 (2004)
  [arXiv:hep-ph/0408120].



\bibitem{superwimp}
  J.~L.~Feng, A.~Rajaraman and F.~Takayama,
  Phys.\ Rev.\ Lett.\  {\bf 91}, 011302 (2003)
  [arXiv:hep-ph/0302215];
 J.~L.~Feng, A.~Rajaraman and F.~Takayama,
  Phys.\ Rev.\ D {\bf 68}, 063504 (2003)
  [arXiv:hep-ph/0306024];
  J.~L.~Feng, S.~Su and F.~Takayama,
  Phys.\ Rev.\ D {\bf 70}, 075019 (2004)
  [arXiv:hep-ph/0404231].

\bibitem{Hooper:2006wv}
  D.~Hooper and A.~M.~Taylor,
  arXiv:hep-ph/0607086.

\bibitem{ilc}
 G.~Weiglein {\it et al.}  [LHC/LC Study Group],
  Phys.\ Rept.\  {\bf 426}, 47 (2006)
  [arXiv:hep-ph/0410364].

\bibitem{baltz}
  E.~A.~Baltz, M.~Battaglia, M.~E.~Peskin and T.~Wizansky,
  arXiv:hep-ph/0602187.

\bibitem{freitas}
  M.~Carena and A.~Freitas,
  arXiv:hep-ph/0608255;
  M.~Carena, A.~Finch, A.~Freitas, C.~Milstene, H.~Nowak and A.~Sopczak,
  Phys.\ Rev.\ D {\bf 72}, 115008 (2005)
  [arXiv:hep-ph/0508152].






\end{thebibliography}
\end{document}